\documentclass[aps,prd,preprint,notitlepage,superscriptaddress,nofootinbib]{revtex4-2}
\usepackage{graphicx}
\usepackage{hyperref}
\usepackage{bm}
\usepackage[normalem]{ulem}
\usepackage{amsmath}
\usepackage{amssymb}
\usepackage[dvipsnames]{xcolor}

\usepackage{lineno}
\newcommand{\eq}[1]{Eq.~\eqref{eq:#1}}

\newcommand{\LDME}[1]{\langle \mathcal{O}^{J/\psi}(#1) \rangle}

\DeclareMathOperator{\sgn}{sgn}

\begin{document}
\global\count\footins=1000 % for footnote overflow

\preprint{MIT-CTP/6079, LA-UR-26-30532, CERN-TH-2026-291}

\title{Quarkonium Energy Correlators in Proton-Proton Collisions:\\
Analytic Calculation and Angular-Momentum Hierarchy}

\date{\today}

\newcommand{\MIT}{Center for Theoretical Physics -- a Leinweber Institute,
  Massachusetts Institute of Technology, Cambridge, MA 02139, USA}
\newcommand{\LANLT}{Theoretical Division, Los Alamos National Laboratory,
  Los Alamos, NM 87545, USA}
\newcommand{\LANLP}{Physics Division, Los Alamos National Laboratory,
  Los Alamos, NM 87545, USA}
\newcommand{\CERN}{Theoretical Physics Department, CERN, 1211 Geneva 23, Switzerland}
\newcommand{\CU}{University of Colorado Boulder, Boulder, CO 80309, USA}

\author{June-Haak Ee}
\email{june-haak.ee@cern.ch}
\affiliation{\MIT}
\affiliation{\LANLT}
\affiliation{\CERN}

\author{Christopher Lee}
\email{clee@lanl.gov}
\affiliation{\LANLT}
\affiliation{\LANLP}

\author{Bianka Me\c caj}
\email{bmecaj@lanl.gov}
\affiliation{\LANLT}

\author{Hugo Pereira Da Costa}
\email{hugo.pereira-da-costa@lanl.gov}
\affiliation{\LANLP}

\author{Bade Sayki}
\email{badesayki@lanl.gov}
\affiliation{\LANLP}
\affiliation{\CU}

%===========================================================================
\begin{abstract}
We present the first complete leading-order (LO) analytic calculation of
the quarkonium energy correlator (QEC) in $pp$ collisions, for the
color-singlet ${}^3S_1^{[1]}$ and the color-octet ${}^1S_0^{[8]}$,
${}^3S_1^{[8]}$, and ${}^3P_J^{[8]}$ channels of $J/\psi$ production. Instead of the high-$p_T$
fragmentation approximation, the calculation retains the full kinematics of
the recoil parton, inverts the map from transverse momentum to the
rest-frame angle in closed form, and is validated by \textsc{HELAC-onia} in shape and
absolute normalization. We also give the analytic counterpart of the
$p_T$-weighted variant of the correlator. The LO shape of each channel is
organized by the total angular momentum $J$ of the $c\bar c$ pair, as a
consequence of the tree-level selection rule that forbids two on-shell
gluons from forming a $J=1$ state. This hierarchy is compressed in the
central sPHENIX acceptance and exposed by the wider forward coverage at the
LHC. 
In the distribution normalized to the total cross section, much though not all of the scale uncertainty cancels, leaving promising distinctions in the QEC shapes between different $J/\psi$ production channels. The discriminating power of the QEC can be further improved by next-to-leading-order calculations and by utilizing QECs in simultaneous fits with the $p_T$ spectrum and polarization measurements.
\end{abstract}
\maketitle
% \clearpage
% \renewcommand{\tocname}{Table of Contents}
% \tableofcontents
% \clearpage

% ===========================================================================
\section{Introduction}
\label{sec:intro}

The production of $J/\psi$ and other quarkonia in hadronic collisions involves both perturbative short-distance scattering and non-perturbative long-distance hadronization, and a consistent description of all data has remained elusive for decades~\cite{Brambilla:2010cs,Andronic:2015wma}.
Within the NRQCD factorization framework~\cite{Bodwin:1994jh}, the inclusive cross section is expanded over a set of intermediate $c\bar{c}$ Fock states $n={}^{2S+1}L_J^{[c]}$ with definite spin ($S$), orbital-angular-momentum ($L$), total-angular-momentum ($J$), and color ($[c]$) quantum numbers, each weighted by a non-perturbative long-distance matrix element (LDME).
The difficulty is that inclusive observables such as the $J/\psi$ transverse-momentum spectrum depend on a linear combination of LDMEs that cannot be individually resolved~\cite{Cho:1995ce,Ma:2010yw,Chao:2012iv}, and the extracted LDME values remain in disagreement between different fits~\cite{Butenschoen:2012px, Gong:2012ug, Bodwin:2014gia, Chung:2022uih}.
More differential observables are therefore needed.
Traditionally, $J/\psi$ polarization has provided a complementary handle on the LDMEs~\cite{STAR:2020igu,CDF:2007msx,CMS:2013gbz,CMS:2024igk}; more recent proposals include $J/\psi$ production within jets~\cite{Baumgart:2014upa,Bain:2016clc,Bain:2016rrv,Bain:2017wvk,Kang:2017yde,Dai:2017cjq,Zhang:2024owr,Wang:2026dul}, which has been measured by the LHCb~\cite{LHCb:2017llq}, CMS~\cite{CMS:2019ebt,CMS:2021puf},  and, in preliminary form, STAR~\cite{Yang:2021smr} experiments.

The quarkonium energy correlator (QEC), proposed in Ref.~\cite{Chen:2024nfl} as a quarkonium-anchored version of the energy-energy correlator~\cite{Basham:1978bw,Dixon:2019uzg,Komiske:2022enw,Craft:2022kdo,Moult:2025nhu}, provides such a probe.
It measures the energy-weighted angular distribution of all final-state particles in the $J/\psi$ rest frame, with the polar axis aligned along the $J/\psi$ flight direction in the lab frame and the polar angle relative to this axis being denoted $\chi$.
At large $J/\psi$ transverse momentum, as reached at high center-of-mass energy, two effects deplete the forward region ($\cos\chi \gtrsim 0$) of perturbative radiation: the recoil parton is boosted strongly backward (toward $\cos\chi \approx -1$) in the $J/\psi$ rest frame, and emission at small angles relative to the $J/\psi$ is suppressed by the dead cone of the heavy quark~\cite{Dokshitzer:1991fd,ALICE:2021aqk}, the mechanism emphasized in Ref.~\cite{Chen:2024nfl}. 
The forward region is thereby made sensitive to the soft non-perturbative gluon emission through which the color-octet $c\bar{c}^{[8]}$ neutralizes its color charge~\cite{Chen:2024nfl}.  Reference~\cite{Chen:2024nfl} computed the perturbative color-singlet contribution at LO for $e^+e^-$ annihilation, with both final-state gluons, and estimated the $pp$ correlator from quarkonium fragmentation functions, leaving out the recoil parton of the hard process.

A subsequent generator-level study at $\sqrt{s}=500\,\mathrm{GeV}$ using \textsc{Pythia\,8}~\cite{Zhang:2026tjo} has explored hadronization effects, finding the hard-parton energy flow backward-peaked, increasingly so as $p_T$ grows, and the forward correlator suppressed by about an order of magnitude from the parton to the hadron level at $p_T>7\,\mathrm{GeV}$; an analytic benchmark against which to compare was not available.
The STAR Collaboration has shown a first, preliminary measurement~\cite{Shen:QM2025} of the $J/\psi$ energy correlator in $pp$ collisions at $\sqrt{s}=500\,\mathrm{GeV}$, for $5<p_T<20\,\mathrm{GeV}$ and $|y|<1$, using a $p_T$-weighted sum over charged hadrons; it had not been published at the time of writing.

The situation changes at lower center-of-mass energies.
At $\sqrt{s}=200\,\mathrm{GeV}$, which is accessible to the sPHENIX experiment at
RHIC~\cite{Belmont:2023fau} and where the inclusive $J/\psi$ spectrum has already been
measured by PHENIX and STAR~\cite{PHENIX:2011gyb,STAR:2012wnc,STAR:2018smh},
the $J/\psi$ Lorentz factor in the partonic center-of-mass frame is of order 1--2 for $p_T\lesssim 5\,\mathrm{GeV}$, where the bulk of the $J/\psi$ cross section lies, so the hard recoil parton populates the backward hemisphere with a much broader $\cos\chi$ distribution than at high $p_T$, where it collapses onto $\cos\chi\approx-1$.
In this regime, a correct analytic treatment of the QEC must retain the full LO matrix elements---not only their high-$p_T$ limit, in which the recoil energy flow collapses onto $\cos\chi\approx-1$---for all partonic subprocesses and NRQCD intermediate states, and must handle the non-trivial kinematics of the $p_T \to \cos\chi$ mapping explicitly.

In this paper we derive the full leading-order (LO) [$\mathcal{O}(\alpha_s^3)$] analytic formula for the QEC, treating all six channels on an equal footing---the color-singlet ${}^3S_1^{[1]}$ and the color-octet production channels leading in $v$ (${}^1S_0^{[8]}$, ${}^3S_1^{[8]}$, ${}^3P_J^{[8]}$ for $J=0,1,2$).
We validate it against \textsc{HELAC-onia} Monte Carlo~\cite{Shao:2012iz,Shao:2015vga} and demonstrate that the LO QEC shape encodes the angular-momentum quantum numbers of the intermediate $c\bar{c}$ Fock state, tracing the pattern to the tree-level two-gluon selection rule that closes the $2\to1$ process $gg\to c\bar c[n]$ for $J=1$ states.
It is known that an open $2\to1$ process controls the small-$p_T$ behavior of quarkonium production~\cite{Petrelli:1997ge,Sun:2012vc}.  Here we show how it orders the $\cos\chi$ shapes of the correlator, identify the acceptances that resolve this ordering, and use the three ${}^3P_J^{[8]}$ substates to show that the ordering depends on $J$ rather than on $L$ or $S$.
This complements the hadronization sensitivity proposed and estimated at high-$p_T$ (LHC) kinematics in Ref.~\cite{Chen:2024nfl}.
We also find that, while the central rapidity acceptance of sPHENIX ($|y|\leq1.1$) is too short in $\cos\chi$ for the six channel shapes to separate, the progressively wider forward coverage of the LHC experiments---ATLAS/CMS \cite{ATLAS:2023qnh, CMS:2017dju} and, most forward, LHCb \cite{LHCb:2015foc}---sharpens the channel-by-channel discrimination and exposes the underlying angular-momentum structure more directly.
Because $\sqrt{s}$ enters the fixed-acceptance prediction only through the parton luminosity, the comparison carries over to the LHC energies at which those wider acceptances are available (Sec.~\ref{subsec:wide_rapidity}).
We additionally provide the analytic counterpart to the $p_{T,i}/M$-weighted variant adopted in Ref.~\cite{Zhang:2026tjo}---which we refer to as the quarkonium $p_T$ correlator (QpTC) for brevity---alongside the QEC in Sec.~\ref{sec:kinematics}, and compare the two observables in Sec.~\ref{subsec:qptc}.

A potential concern is that NRQCD predictions at leading order in $\alpha_s$ are known to be subject to large higher-order corrections in inclusive observables, particularly at high $J/\psi$ transverse momentum where the LDME values extracted at LO and NLO can differ substantially~\cite{Butenschoen:2012qr,Chao:2012iv,Chung:2022uih}. Two considerations motivate the LO treatment in the present context, and one limits it. 
First, although the inclusive $J/\psi$ $p_T$ spectrum is known at NLO in NRQCD~\cite{Butenschoen:2012qr,Chao:2012iv}, the QEC is a distinct, energy-weighted differential observable whose NLO extension---the one-loop $2\to2$ and real $2\to3$ corrections to the correlator itself---has not been carried out; our analytic LO result supplies the recoil-parton contribution to the correlator, computed without expanding in $M/p_T$, and the baseline on which higher-order and Monte Carlo work can build.
Second, given that currently available MC event generators such as \textsc{Pythia\,8}~\cite{Sjostrand:2014zea,Bierlich:2022pfr} implement only the LO hard matrix elements for these quarkonia production processes, supplemented in recent developments by quarkonium production through NRQCD splittings in the timelike parton shower~\cite{Cooke:2023ldt,Masouminia:2025kec}, a clear understanding of the LO analytic baseline provides a clean reference against which the corrections from parton showering, multi-parton interactions, and hadronization can eventually be studied separately.
The limitation to our LO analysis is kinematic: the correlator is dominated by low transverse momenta, with half of the integrated correlator from $p_T\lesssim1.4\,\mathrm{GeV}$ at sPHENIX and $p_T\lesssim2.2\,\mathrm{GeV}$ at the LHC acceptances and about $90\%$ or more from $p_T<7\,\mathrm{GeV}$---below the range in which the LDMEs of Table~\ref{tab:ldme} were fitted, and where fixed-order NRQCD must be supplemented by soft-gluon resummation~\cite{Chao:2012iv,Sun:2012vc}. 
We therefore use the LO result as a baseline for the shape rather than as a precise quantitative prediction, and return to this in Sec.~\ref{subsec:pt_tuning}.
The order at which each ingredient of the calculation enters---matrix elements, parton densities, and the LDME fit---is collected in Sec.~\ref{subsec:numerical_inputs}.

The paper is organized as follows.
Section~\ref{sec:nrqcd} reviews the NRQCD framework and defines the QEC and QpTC.
Section~\ref{sec:kinematics} presents the kinematics and derives the analytic QEC and QpTC formulas.
Section~\ref{sec:amplitudes} collects the short-distance amplitudes, lists the numerical inputs, validates the analytic result against \textsc{HELAC-onia}, compares the QEC with the QpTC of
Ref.~\cite{Zhang:2026tjo}, and presents the LDME-weighted analytic baseline predictions that frame the rest of the paper.
Section~\ref{sec:soft} quantifies the soft-gluon contribution from color-octet hadronization.
Section~\ref{sec:interpretation} compares the channel shapes at the sPHENIX, ATLAS/CMS and LHCb acceptances, showing how widening the rapidity acceptance sharpens the channel-by-channel discrimination, traces the grouping of the shapes to the total angular momentum of the intermediate $c\bar{c}$ state, and examines the dependence on the collinear cutoff.
Section~\ref{sec:ldme} assesses the leading-order LDME-discriminating power of the QEC and QpTC across three distinctive NRQCD scenarios and discusses the sPHENIX and LHC prospects.
Section~\ref{sec:conclusion} concludes.
Appendix~\ref{app:ptsol} provides the closed-form inversion of $\cos\chi(p_T)$, and Appendix~\ref{app:coschi_max} derives the forward kinematic boundary of $\cos\chi$ within the fiducial acceptance.

% ===========================================================================
\section{NRQCD Framework and the Quarkonium Energy Correlator}
\label{sec:nrqcd}

\subsection{NRQCD factorization}
We adopt the NRQCD factorization framework of Ref.~\cite{Bodwin:1994jh}, in which the $J/\psi$ production cross section is written as
\begin{equation}
\label{eq:nrqcd_factorization}
\sigma(pp \to J/\psi + X) =
\sum_n \hat{\sigma}(pp \to c\bar{c}[n] + X)\,
\LDME{n},
\end{equation}
where $\hat{\sigma}$ is the perturbatively calculable short-distance cross section for producing a $c\bar{c}$ pair in the quantum state $n$, and $\LDME{n}$ is the LDME for its non-perturbative transition into the physical $J/\psi$.
At leading order in the velocity $v$, the relevant NRQCD channels are
\begin{equation}
\label{eq:channels}
n \;=\; {}^3S_1^{[1]},\quad {}^1S_0^{[8]},\quad {}^3S_1^{[8]},\quad {}^3P_J^{[8]}
\quad (J=0,1,2),
\end{equation}
where $v$ is the typical relative velocity of the $c$ and $\bar{c}$ in the quarkonium rest frame ($v^2 \approx 0.3$ for charmonium), the small parameter in which the NRQCD expansion is organized alongside $\alpha_s$.
The color-singlet LDME scales as $v^3$ and is not fitted to production data but computed from the wave function at the origin of a potential model (Sec.~\ref{subsec:numerical_inputs}), while the color-octet LDMEs scale as $v^7$ and must be extracted from data~\cite{Cho:1995vh,Cho:1995ce,Chung:2022uih}.
Heavy-quark spin symmetry, exact at leading order in $v$, relates the three $P$-wave matrix elements, $\LDME{{}^3P_J^{[8]}}=(2J+1)\LDME{{}^3P_0^{[8]}}$ for $J=0,1,2$~\cite{Bodwin:1994jh}, so the six channels of Eq.~\eqref{eq:channels} are governed by three independent color-octet LDMEs---$\LDME{{}^1S_0^{[8]}}$, $\LDME{{}^3S_1^{[8]}}$, and $\LDME{{}^3P_0^{[8]}}$---in addition to the fixed color-singlet one.  The three ${}^3P_J^{[8]}$ substates therefore enter every fit only through their fixed linear combination; 
we nonetheless resolve them individually in Secs.~\ref{sec:amplitudes}--\ref{sec:interpretation}, because the three substates share $L=1$ and $S=1$ and differ only in $J$, so that the differences among their $\cos\chi$ shapes can be attributed to $J$ alone.

While the color-singlet ${}^3S_1^{[1]}$ channel forms the physical $J/\psi$ directly (the produced $c\bar{c}$ already has $J/\psi$ quantum numbers), the color-octet $c\bar{c}^{[8]}$ intermediate states must emit soft gluons before forming the color-singlet $J/\psi$. In the NRQCD velocity expansion, the ${}^1S_0^{[8]}$ and ${}^3P_J^{[8]}$ states each reach a color-singlet $c\bar{c}$ configuration with $J/\psi$ quantum numbers through a single soft-gluon transition: a chromomagnetic (M1) transition for ${}^1S_0^{[8]}$, whose spin must flip from $S=0$ to $S=1$, and a chromoelectric (E1) transition for ${}^3P_J^{[8]}$, whose orbital angular momentum changes by one unit. 
The ${}^3S_1^{[8]}$ state can reach a color-singlet $c\bar{c}$ configuration with $J/\psi$ quantum numbers either directly through two chromoelectric (E1) transitions, or indirectly through hadronization into an intermediate $\chi_{cJ}$ ($= {}^3P_J^{[1]}$) state followed by the radiative decay $\chi_{cJ}\to J/\psi+\gamma$; the inclusive LDME $\LDME{{}^3S_1^{[8]}}$ sums over all such hadronization paths contributing to the inclusive $J/\psi$~\cite{Cho:1995vh,Cho:1995ce}.
The feed-down enters our calculation only through this normalization: the $\chi_{cJ}\to J/\psi+\gamma$ (and $\psi(2S)\to J/\psi+X$) decays themselves are not simulated, and their effect on the energy flow is discussed in Sec.~\ref{subsec:ldme_implications}.

\subsection{Definition of the quarkonium energy correlator and its $p_T$-weighted variant}

Following Ref.~\cite{Chen:2024nfl}, the QEC is defined as the energy-weighted differential cross section
\begin{equation}
\label{eq:qec_def}
\frac{d\Sigma}{d\cos\chi}
\;=\;
\sum_X
\int d\sigma(pp \to J/\psi + X)
\sum_{i\in X}
\frac{E_i}{M}
\;\delta\!\left(\cos\chi - \cos\chi_i\right),
\end{equation}
where $X$ denotes the set of all final-state particles other than the $J/\psi$, and $E_i$, $\chi_i$ are the energy and polar angle of particle $i \in X$, both evaluated in the $J/\psi$ rest frame. The polar axis is defined by boosting from the $pp$ center-of-mass (lab) frame to the $J/\psi$ rest frame along the lab-frame $J/\psi$ three-momentum, so that in the rest frame the $\hat{z}$-axis points along the original $J/\psi$ flight direction (helicity frame~\cite{Faccioli:2010kd}). We refer to the hemisphere $\cos\chi > 0$ (emission aligned with the $J/\psi$ flight direction) as forward and to $\cos\chi < 0$ (emission toward the recoiling parton) as backward. The energy weighting by $E_i/M$ ensures infrared and collinear safety, and following Ref.~\cite{Chen:2024nfl} the energy is normalized to the fixed quarkonium mass $M$, so that $E_i/M$ is a dimensionless energy fraction. The rest-frame recoil energy $E_i$ itself varies event by event with the partonic kinematics [at LO, $E_i = (\hat{s}-M^2)/(2M)$; see Eq.~\eqref{eq:E4pp_shat}].

We will also consider a transverse-momentum-based variant of the QEC, in which the rest-frame energy weighting $E_i/M$ is replaced by the lab-frame transverse-momentum weighting $p_{T,i}/M$.  Transverse-momentum weighting is standard for energy correlators at hadron colliders~\cite{Komiske:2022enw,Chen:2020vvp}; we adopt the $p_{T,i}/M$ weighting of the recent generator-level \textsc{Pythia\,8} study of Ref.~\cite{Zhang:2026tjo}, but keep the other conventions of Eq.~\eqref{eq:qec_def}: a cross section summed over all final-state particles. Reference~\cite{Zhang:2026tjo} normalizes per $J/\psi$ instead, which changes only the overall normalization, and at the hadron level sums over charged hadrons only.
We refer to this observable as the quarkonium $p_T$ correlator (QpTC):
\begin{equation}
\label{eq:qptc_def}
\frac{d\Sigma^{\rm QpTC}}{d\cos\chi}
\;=\;
\sum_X
\int d\sigma(pp \to J/\psi + X)
\sum_{i\in X}
\frac{p_{T,i}}{M}
\;\delta\!\left(\cos\chi - \cos\chi_i\right),
\end{equation}
where $p_{T,i}$ is the lab-frame transverse momentum of particle $i$ and $\chi_i$ is the same rest-frame polar angle as in the QEC definition. Reference~\cite{Zhang:2026tjo} reported a fully showered and hadronized \textsc{Pythia\,8} prediction for the QpTC at $\sqrt{s}=500\,\mathrm{GeV}$ but did not provide an analytic counterpart; we develop the analytic QpTC alongside the QEC in Sec.~\ref{subsec:analytic_QEC} and compare the two observables in Sec.~\ref{subsec:qptc}. The two weightings are equally accessible in the analytic calculation but they differ in experimental character: $p_{T,i}$ is invariant under longitudinal boosts and down-weights the high-lab-rapidity region populated by beam remnants and the underlying event, where detector coverage is poorest, whereas the rest-frame energy weight enhances that region. The price is a $\cos\chi$-dependent dilution of the hard signal, quantified in Sec.~\ref{subsec:qptc}.

The leading partonic process supplying a hard recoil is the $2\to 2$ production $ab\to c\bar{c}[n]+k$, which enters at $\mathcal{O}(\alpha_s^3)$ for both color-singlet and color-octet channels; here $a,b$ are the incoming partons and $k$ the recoil parton.  At this order the final state $X$ consists of that single recoil parton, and the QEC receives contributions from it alone.  The $2\to1$ processes available at $\mathcal{O}(\alpha_s^2)$ produce no recoil parton and would contribute to Eq.~\eqref{eq:qec_def} only through the soft gluon of the octet-to-singlet transition; at this order they carry $J/\psi$ transverse momentum of the order of that soft-gluon energy, $E_g\approx0.2\,\mathrm{GeV}$ in the model of Sec.~\ref{sec:soft}, and are removed by the fiducial requirement $p_T\geq p_T^{\min}$ of Sec.~\ref{subsec:analytic_QEC}. 
Beyond this order, initial-state radiation spreads these events over $p_T\lesssim M$~\cite{Sun:2012vc}, the region
whose treatment we discuss in Sec.~\ref{subsec:pt_tuning}.
In the color-octet channels, the soft gluon emitted in hadronization adds a second particle to $X$, which we estimate in Sec.~\ref{sec:soft}; the full multi-particle final state generated by parton showering, multi-parton interactions, and hadronization lies beyond the scope of the present analytic study.

% ===========================================================================
\section{Kinematics and Analytic QEC Formula}
\label{sec:kinematics}

\subsection{Partonic kinematics and hadronic cross section}

\begin{figure}
\centering
\includegraphics[width=\linewidth]{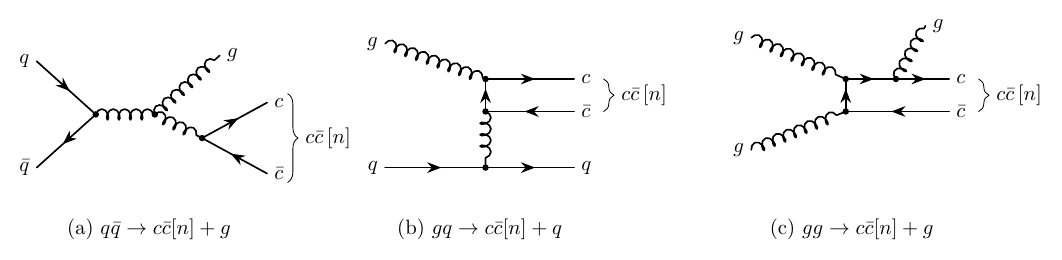}
\caption{Representative leading-order [$\mathcal{O}(\alpha_s^3)$]
  Feynman diagrams for the $2\to 2$ partonic subprocesses producing a
  $c\bar{c}[n]$ pair with a recoil parton:
  (a)~$q\bar{q}\to c\bar{c}[n]+g$ via $s$-channel gluon annihilation,
  (b)~$gq\to c\bar{c}[n]+q$ via $t$-channel gluon exchange, and
  (c)~$gg\to c\bar{c}[n]+g$. One representative diagram
  is shown per subprocess; the complete sets follow
  Refs.~\cite{Baier:1983va,Gastmans:1987be} for the color singlet and
  Ref.~\cite{Cho:1995ce} (Fig.~3 therein) for the color octet.}
\label{fig:feynman}
\end{figure}

The leading-order $2\to 2$ processes contributing to $J/\psi$ production
with nonzero transverse momentum are
\begin{align}
\label{eq:processes}
q\bar{q} &\;\to\; c\bar{c}[n] + g,
\nonumber \\
gq &\;\to\; c\bar{c}[n] + q,
\nonumber \\
gg &\;\to\; c\bar{c}[n] + g,
\end{align}
for each NRQCD state $n$; representative diagrams are shown in Fig.~\ref{fig:feynman}.
Throughout, we refer to the NRQCD states $n$ as channels, to the partonic initial states $q\bar q$, $gq$ and $gg$ as subprocesses, and to $2\to1$ and $2\to2$ as processes.

We parameterize the kinematics with light-like vectors
\begin{equation}
\label{eq:lightcone}
n^\mu = (1,0,0,1), \qquad \bar{n}^\mu = (1,0,0,-1),
\end{equation}
in the $pp$ CM frame, with $n^2=\bar{n}^2=0$ and $n\cdot\bar{n}=2$.
Let us take the $pp$ CM energy to be $\sqrt{s}$.  The incoming hadron momenta are then, neglecting the proton mass,
\begin{equation}
P_1^\mu = \sqrt{s}\,\frac{n^\mu}{2},\qquad
P_2^\mu = \sqrt{s}\,\frac{\bar{n}^\mu}{2},
\end{equation}
with $(P_1+P_2)^2 = s$.  The incoming partons carry momentum fractions $x_1,\,x_2$ of their parent hadrons,
\begin{equation}
p_1^\mu = x_1\sqrt{s}\,\frac{n^\mu}{2},\qquad
p_2^\mu = x_2\sqrt{s}\,\frac{\bar{n}^\mu}{2}.
\end{equation}
The final-state $J/\psi$ and recoil momenta are
\begin{align}
\label{eq:final_momenta}
p_3^\mu &= E_T\,e^{y_3}\,\frac{n^\mu}{2}
          + E_T\,e^{-y_3}\,\frac{\bar{n}^\mu}{2}
          + p_T\, \hat{n}_T^\mu,
\nonumber \\
p_4^\mu &= p_T\,e^{y_4}\,\frac{n^\mu}{2}
          + p_T\,e^{-y_4}\,\frac{\bar{n}^\mu}{2}
          - p_T\, \hat{n}_T^\mu,
\end{align}
where $y_3$, $y_4$ are the $J/\psi$ and recoil rapidities, respectively, $p_T$ is the common transverse momentum magnitude, and $\hat{n}_T^\mu = (0,\bm{\hat{n}}_T,0)$ is a unit transverse vector (its azimuthal orientation is irrelevant after integration); we use $E_T \equiv \sqrt{M^2+p_T^2}$ throughout. The on-shell conditions $p_3^2=M^2$ and $p_4^2=0$ are satisfied by construction.
The parton momentum fractions are fixed by energy-momentum conservation:
\begin{align}
\label{eq:xfractions}
x_1 &= \frac{1}{\sqrt{s}}\left(E_T\,e^{y_3} + p_T\,e^{y_4}\right),
\nonumber \\
x_2 &= \frac{1}{\sqrt{s}}\left(E_T\,e^{-y_3} + p_T\,e^{-y_4}\right).
\end{align}
The partonic Mandelstam variables satisfy
$\hat{s}+\hat{t}+\hat{u}=M^2$ and are given explicitly by
\begin{align}
\label{eq:mandelstam}
\hat{s} &= M^2 + 2p_T^2 + 2p_T E_T\cosh\Delta_y,
\nonumber \\
\hat{t} &= -p_T E_T\,e^{-\Delta_y} - p_T^2,
\nonumber \\
\hat{u} &= -p_T E_T\,e^{+\Delta_y} - p_T^2,
\end{align}
where we introduce the rapidity difference $\Delta_y \equiv y_3-y_4$.
The triple-differential hadronic cross section is then \cite{Cho:1995ce,Cho:1995vh}
\begin{equation}
\label{eq:hadronic_xsec}
\frac{d^3\sigma}{dy_3\,dy_4\,dp_T}
= 2p_T \sum_{a,b,k}\sum_n
  x_1 x_2\,
  f_{a/p}(x_1,\mu_F)\,f_{b/p}(x_2,\mu_F)\,
  \frac{d\hat{\sigma}}{d\hat{t}}(ab\to c\bar{c}[n]+k)\,
  \LDME{n},
\end{equation}
where $f_{a/p}(x,\mu_F)$ is the parton distribution function for a parton $a$ from a proton $p$ with the longitudinal momentum fraction $x$ at the factorization scale $\mu_F$.
The partonic cross section $d\hat{\sigma}/d\hat{t}$ is defined by
\begin{equation}
\label{eq:partonic_xsec}
\frac{d\hat{\sigma}}{d\hat{t}}(ab\to c\bar{c}[n]+k)
= \frac{1}{16\pi\hat{s}^2}
  \;\overline{\sum}\,|\mathcal{A}(ab\to c\bar{c}[n]+k)|^2,
\end{equation}
where $\overline{\sum}$ denotes averaging over initial-state spins and colors and summing over final-state spins and colors. The partonic short-distance amplitude $\mathcal{A}(ab\to c\bar{c}[n]+k)$ is independent of the long-distance hadronization encoded by $\LDME{n}$, and the sum over $n$ in Eq.~\eqref{eq:hadronic_xsec} runs over the NRQCD intermediate channels listed in Eq.~\eqref{eq:channels}.
In general the partonic cross section depends on both $\mu_R$ and $\mu_F$, the latter through the collinear counterterms that compensate the scale dependence of the parton distributions; at the leading order considered here it depends only on $\mu_R$, through $\alpha_s^3(\mu_R)$, and the $\mu_F$ dependence of the parton distributions is left uncompensated (Sec.~\ref{subsec:wide_rapidity}).  The scale choices are given in Sec.~\ref{subsec:numerical_inputs}.

\subsection{The polar angle in the \texorpdfstring{$J/\psi$}{J/psi} rest frame}

To compute the QEC we express the recoil momentum $p_4$ in the $J/\psi$ rest frame via the combined transformation $\Lambda = \Lambda_{\rm boost}\,\Lambda_{\rm rot}$, where $\Lambda_{\rm rot}$ aligns the $J/\psi$ three-momentum with $+\hat{z}$ and $\Lambda_{\rm boost}$ boosts to rest.

Introducing the $J/\psi$ three-momentum magnitude
\begin{equation}
\label{eq:p3mag}
|\vec{p}_3| = \sqrt{E_T^2\sinh^2 y_3 + p_T^2},
\end{equation}
the rotation matrix in the $xz$-plane (assuming $\hat{n}_T^\mu$ is along the $x$ axis) is
\begin{equation}
\Lambda_{\rm rot} =
\begin{pmatrix}
1 & 0 & 0 & 0 \\
0 & \cos\theta & 0 & -\sin\theta \\
0 & 0 & 1 & 0 \\
0 & \sin\theta & 0 & \cos\theta
\end{pmatrix},
\quad
\cos\theta = \frac{E_T\sinh y_3}{|\vec{p}_3|},\quad
\sin\theta = \frac{p_T}{|\vec{p}_3|}.
\end{equation}
The subsequent boost along $-z$ that brings $p_3$ to rest is
\begin{equation}
\Lambda_{\rm boost} =
\begin{pmatrix}
\gamma & 0 & 0 & -\beta\gamma \\
0 & 1 & 0 & 0 \\
0 & 0 & 1 & 0 \\
-\beta\gamma & 0 & 0 & \gamma
\end{pmatrix},
\quad
\gamma = \frac{E_T\cosh y_3}{M},
\quad
\beta\gamma = \frac{|\vec{p}_3|}{M},
\end{equation}
so that $\Lambda_{\rm boost}$ acting on the rotated $J/\psi$ momentum $(E_T\cosh y_3, 0, 0, |\vec{p}_3|)$ yields $(M, 0, 0, 0)$.
After the combined transformation $\Lambda = \Lambda_{\rm boost}\Lambda_{\rm rot}$, the recoil momentum components in the $J/\psi$ rest frame are
\begin{align}
\label{eq:p4_rest}
E_4^{\rm rest} &= \frac{p_T(p_T+E_T\cosh\Delta_y)}{M}, \nonumber\\
p_{4x}^{\rm rest} &= -\frac{p_T(E_T\sinh y_3 + p_T\sinh y_4)}{|\vec{p}_3|}, \nonumber\\
p_{4y}^{\rm rest} &= 0, \nonumber\\
p_{4z}^{\rm rest} &= -\frac{p_T[p_T\cosh y_3(E_T+p_T\cosh\Delta_y)
            + M^2\sinh y_3\sinh\Delta_y]}
           {M\,|\vec{p}_3|}.
\end{align}
One verifies that $(E_4^{\rm rest})^2 - (p_{4x}^{\rm rest})^2 - (p_{4y}^{\rm rest})^2 - (p_{4z}^{\rm rest})^2 = 0$.
We can further express the recoil's rest-frame energy directly in terms of the partonic Mandelstam invariant $\hat{s}$,
\begin{equation}
\label{eq:E4pp_shat}
E_4^{\rm rest} = \frac{\hat{s}-M^2}{2M},
\end{equation}
a two-body kinematic identity that exhibits the QEC weight $E_4^{\rm rest}/M = (\hat{s}-M^2)/(2M^2)$ as a direct probe of the partonic center-of-mass energy.
Consequently, the cosine of the polar angle $\chi$ defined in Eq.~\eqref{eq:qec_def} is $\cos\chi = p_{4z}^{\rm rest}/E_4^{\rm rest}$, which evaluates to
\begin{align}
\label{eq:coschi}
\cos\chi(p_T; y_3, y_4)
&= -
  \frac{p_T\cosh y_3
    \bigl(E_T+p_T\cosh\Delta_y\bigr)
    + M^2\sinh y_3\sinh\Delta_y}
  {\bigl(p_T+E_T\cosh\Delta_y\bigr)\,|\vec{p}_3|}.
\end{align}

\begin{figure}
\centering
\includegraphics[width=0.65\linewidth]{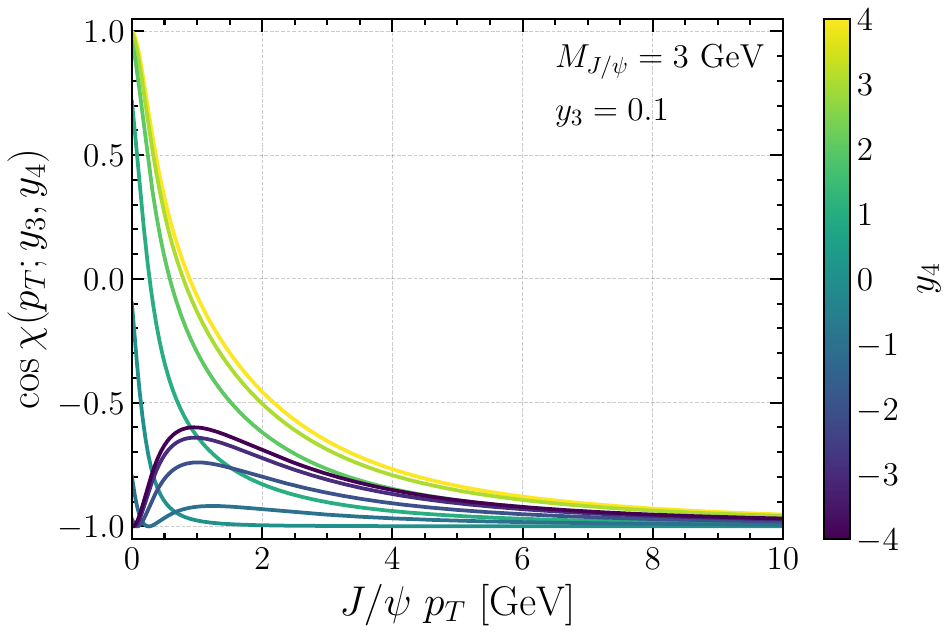}
\caption{$\cos\chi$ of Eq.~\eqref{eq:coschi} as a function of $p_T$ for $y_3=0.1$ and nine recoil rapidities
  $y_4=-4,\dots,+4$ (color scale), with $M=3.0\,\mathrm{GeV}$.  Each curve starts at $-\sgn(y_3)\tanh\Delta_y$ and
  approaches $-1$ at large $p_T$; for $y_4\lesssim-1$ it first passes through a local maximum, the non-monotonic map
  discussed in the text.}
\label{fig:coschi_pt}
\end{figure}

A key property of Eq.~\eqref{eq:coschi} is that the map $p_T \mapsto \cos\chi$ is non-monotonic at fixed $(y_3,y_4)$: a given value of $\cos\chi$ can correspond to multiple values of $p_T$.
This is illustrated in Fig.~\ref{fig:coschi_pt}: at fixed $(y_3, y_4)$ the distribution can display local extrema in the backward region before approaching $\cos\chi\to -1$ at large $p_T$.
Physically, at $p_T\to 0$ the recoil rest-frame angle is fixed by the rapidity difference, $\cos\chi \to -\sgn(y_3)\tanh(y_3-y_4)$, while at $p_T\to\infty$ the $J/\psi$ Lorentz boost dominates and $\cos\chi\to -1$ irrespective of rapidity. More generally, the departure of $\cos\chi$ from $-1$ is a quarkonium-mass effect, generically of $\mathcal{O}(M^2/p_T^2)$, vanishing at large $p_T$ or, equivalently, in the massless limit: as $M\to0$, $E_T\to p_T$ and $|\vec p_3|\to p_T\cosh y_3$, the $M^2$ term in the numerator of Eq.~\eqref{eq:coschi} drops out, numerator and denominator reduce to the same expression, and $\cos\chi\to-1$ identically for any $(p_T,y_3,y_4)$---a massless quarkonium and its recoil are exactly back-to-back. 
A leading-power fragmentation description keeps only the first term of this expansion, and there the recoil-parton energy flow sits entirely at $\cos\chi=-1$; its whole $\cos\chi$ dependence is a power correction.

For intermediate $p_T$, where this mass effect is $\mathcal{O}(1)$, the competition between the partonic emission angle and the rest-frame boost can produce a local extremum, making the map $p_T\mapsto\cos\chi$ non-monotonic; this happens only when the recoil lies on the opposite side of the $J/\psi$ and farther from mid-rapidity ($y_4\lesssim-1$ in Fig.~\ref{fig:coschi_pt}), so a same-side forward acceptance such as LHCb's has a single root throughout (Sec.~\ref{subsec:helac}).
This non-monotonic structure is a feature of the recoil-parton energy flow---the contribution that Ref.~\cite{Chen:2024nfl} sets aside in its illustrative estimate, on the grounds that it typically contributes to $\cos\chi<0$.  We confirm that it is indeed concentrated at $\cos\chi<0$, but find that with a suitably chosen acceptance it reaches into $\cos\chi>0$ as well (Sec.~\ref{sec:interpretation}).

\subsection{Analytic QEC formula}
\label{subsec:analytic_QEC}

Inserting the $2\to 2$ hadronic cross section into Eq.~\eqref{eq:qec_def} and trading the $p_T$ integration for a $\cos\chi$ integration via the delta function, we obtain
\begin{equation}
\label{eq:qec_main}
\frac{d^3\Sigma}{dy_3\,dy_4\,d\cos\chi}
= \int_0^\infty dp_T\;
  \delta\bigl[\cos\chi - \cos\chi(p_T; y_3, y_4)\bigr]\;
  \frac{E_4^{\rm rest}}{M}\;
  \frac{d^3\sigma}{dy_3\,dy_4\,dp_T}\,,
\end{equation}
where $\cos\chi(p_T;y_3,y_4)$ is given by the formula in \eq{coschi}.
Evaluating the delta function yields our central result:
\begin{equation}
\label{eq:qec_analytic}
\frac{d^3\Sigma}{dy_3\,dy_4\,d\cos\chi}
= \sum_{p_T = p_T^{\rm sol}(\cos\chi)}
  \left|\frac{d\cos\chi}{dp_T}\right|^{-1}
  \cdot
  \frac{p_T\!\left(p_T + E_T\cosh\Delta_y\right)}{M^2}
  \cdot
  \frac{d^3\sigma}{dy_3\,dy_4\,dp_T},
\end{equation}
where the sum runs over all roots $p_T^{\rm sol}$ of $\cos\chi(p_T) = \cos\chi$, and the Jacobian factor $|d\cos\chi/dp_T|^{-1}$ follows from the analytic derivative of Eq.~\eqref{eq:coschi},
\begin{equation}
\label{eq:jacobian}
\frac{d\cos\chi}{dp_T}
= \frac{\mathcal{N}'\mathcal{D} - \mathcal{N}\mathcal{D}'}{\mathcal{D}^2},
\end{equation}
where $\mathcal{N}$ and $\mathcal{D}$ are the numerator and denominator of Eq.~\eqref{eq:coschi}, respectively, and $\mathcal{N}'$ and $\mathcal{D}'$ are their derivatives in $p_T$. 
At a turning point of the non-monotonic map, where $d\cos\chi/dp_T=0$, the Jacobian factor diverges and the analytic QEC develops an integrable square-root singularity.

The $p_T$ solutions in Eq.~\eqref{eq:qec_analytic} are obtained in closed form.  Equation~\eqref{eq:coschi} involves only the two radicals $E_T=\sqrt{p_T^2+M^2}$ and $|\vec p_3|$ [Eq.~\eqref{eq:p3mag}], each with a radicand polynomial in $p_T^2$; isolating and squaring them in turn removes both radicals and reduces the constraint $\cos\chi(p_T)=\cos\chi$ to a fourth-order polynomial equation in $p_T^2$,
\begin{equation}
\label{eq:quartic}
\sum_{k=0}^{4} a_k\,p_T^{2k} = 0 ,
\end{equation}
whose coefficients $a_k$ are elementary functions of $(y_3,y_4,\cos\chi)$ and $M$.  Its four roots follow from Ferrari's method~\cite{Abramowitz:1964} (Appendix~\ref{app:ptsol}); the physical $p_T^{\rm sol}$ are the real, positive solutions that reproduce the input $\cos\chi$ on back-substitution, squaring having introduced spurious roots.  Over the fiducial acceptance a single physical root usually survives, with up to three where the non-monotonic $p_T\mapsto\cos\chi$ map folds; the sum in Eq.~\eqref{eq:qec_analytic} runs over all such roots.
The analytic QpTC follows from the identical manipulation with the QEC weight $E_4^{\rm rest}/M$ of Eq.~\eqref{eq:qec_main} replaced by the lab-frame weight $p_T/M$ of its definition [Eq.~\eqref{eq:qptc_def}],
\begin{equation}
\label{eq:qptc_analytic}
\frac{d^3\Sigma^{\rm QpTC}}{dy_3\,dy_4\,d\cos\chi}
= \sum_{p_T = p_T^{\rm sol}(\cos\chi)}
  \left|\frac{d\cos\chi}{dp_T}\right|^{-1}
  \cdot \frac{p_T}{M}
  \cdot \frac{d^3\sigma}{dy_3\,dy_4\,dp_T},
\end{equation}
with the same set of $p_T$ roots and Jacobian as the QEC.
The physical QEC and QpTC are obtained by integrating the respective differential distributions over the experimental acceptance:
\begin{equation}
\label{eq:qec_integrated}
\frac{d\Sigma}{d\cos\chi}
= \int_{y_3^{\min}}^{y_3^{\max}} dy_3\!
  \int_{y_4^{\min}}^{y_4^{\max}} dy_4\;
  \frac{d^3\Sigma}{dy_3\,dy_4\,d\cos\chi},
\end{equation}
where $d^3\Sigma/(dy_3\,dy_4\,d\cos\chi)$ stands for either the QEC [Eq.~\eqref{eq:qec_analytic}] or the QpTC [Eq.~\eqref{eq:qptc_analytic}] differential distribution.

Throughout this work we adopt a fiducial definition of the QEC that matches the central-rapidity acceptance of the sPHENIX detector at RHIC~\cite{Belmont:2023fau}.   At LO the final state consists of the $J/\psi$ and a single recoil parton, and we require both to lie within the central acceptance,
\begin{equation}
\label{eq:event_cuts}
|y_3| \leq 1.1, \qquad |y_4| \leq 1.1, \qquad p_T \geq p_T^{\min},
\end{equation}
with $p_T^{\min} = 0.5\,\mathrm{GeV}$.  Here $|y_3| \leq 1.1$ places the
$J/\psi$ within the sPHENIX barrel acceptance;
$|y_4| \leq 1.1$ requires the recoil parton---effectively massless at LO,
so $|\eta_4| = |y_4|$---to lie in the same acceptance and fixes the
range of the $y_4$ integration in Eq.~\eqref{eq:qec_integrated}; 
and the transverse-momentum threshold $p_T^{\min}$ is the hard-scattering (collinear) cutoff that avoids the singularity of the LO matrix elements as $p_T \to 0$; it is moreover experimentally motivated, since a recoil parton at vanishing $p_T$---soft, or collinear with the
beam---escapes reconstruction.\footnote{The cutoff regulates that singularity; the limits of fixed-order collinear factorization at $p_T\lesssim M$, and the sensitivity of our results to $p_T^{\min}$, are discussed in Sec.~\ref{subsec:pt_tuning}.}
Since $p_{T,4} = p_T$ at LO, no separate cut on
the recoil transverse momentum is needed.  These fiducial cuts are
applied consistently to the analytic QEC and QpTC [Eq.~\eqref{eq:qec_integrated}]
and to the \textsc{HELAC-onia} validation of Sec.~\ref{sec:amplitudes}.
The extension to wider rapidity acceptances, and the dependence on $p_T^{\min}$, are discussed in Sec.~\ref{sec:interpretation}.

% ===========================================================================
\section{Analytic Results and Validation against \textsc{HELAC-onia}}
\label{sec:amplitudes}

\subsection{Squared amplitudes}

The partonic short-distance squared amplitudes $\overline{\sum}|\mathcal{A}|^2$ are taken from Refs.~\cite{Baier:1983va,Gastmans:1987be} (color-singlet) and Refs.~\cite{Cho:1995ce,Cho:1995vh} (color-octet).
The color-octet amplitudes of Refs.~\cite{Cho:1995ce,Cho:1995vh} are given as helicity-resolved expressions for the intermediate $c\bar{c}$ pair, with the helicity quantization axis defined along the pair's flight direction in the partonic CM frame. We sum over the $c\bar{c}$ helicities to obtain the spin-averaged short-distance amplitude that, combined with the LDMEs $\LDME{n}$ via Eqs.~\eqref{eq:hadronic_xsec}--\eqref{eq:partonic_xsec}, yields the unpolarized cross section relevant for the QEC.
The color-singlet ${}^3S_1^{[1]}$ channel contributes to $J/\psi$ production at $\mathcal{O}(\alpha_s^3)$ only through gluon-gluon fusion: the $q\bar{q}$ and $gq$ initial states yield no LO color-singlet contribution due to the negative $C$-parity of the $J/\psi$ and the color structure of single-gluon exchange. The sums over $q$ in the $q\bar{q}$ and $gq$ contributions run over the three light flavors $u, d, s$, as in the \textsc{HELAC-onia}~\cite{Shao:2012iz,Shao:2015vga} samples used for the validation in Sec.~\ref{subsec:helac}; charm- and bottom-quark initial states are omitted in both: the charm density at the relevant momentum fractions is 4--6\% (RHIC) to 12--17\% (LHC) of the light-quark one, so it would add a few percent to the total, far below the scale uncertainty.

\subsubsection{\texorpdfstring{$q\bar{q}\to J/\psi+g$}{qqbar to J/psi+g}}

In the $q\bar{q}$ subprocess [Fig.~\ref{fig:feynman}(a)],
the $c\bar{c}$ formation proceeds through $s$-channel virtual-gluon
annihilation accompanied by an additional outgoing gluon, yielding
squared amplitudes whose denominators are powers of $\hat s$ and $\hat s-M^2$, except for the
helicity-$|h|=1$ ${}^3S_1^{[8]}$ term, which carries a $1/(\hat t\hat u)$ pole from radiation off the
incoming quarks. As a representative example,
the short-distance squared amplitude for $c\bar{c}[{}^1S_0^{[8]}]$
formation in this channel is
\begin{equation}
\label{eq:amp_qq_1S08}
\overline{\sum}|\mathcal{A}(q\bar{q}\to c\bar{c}[{}^1S_0^{[8]}]\,g)|^2
= \frac{5(4\pi\alpha_s)^3}{27\,M}
  \frac{\hat{t}^2+\hat{u}^2}{\hat{s}\,(\hat{s}-M^2)^2},
\end{equation}
which corresponds to Eq.~(A3a) of Ref.~\cite{Cho:1995ce}; being a
spin-0 state, the ${}^1S_0^{[8]}$ requires no helicity sum. The
expressions for ${}^3S_1^{[8]}$ and ${}^3P_J^{[8]}$ in this channel
are given in Eqs.~(A3b)--(A3i) of the same reference.

\subsubsection{\texorpdfstring{$gq\to J/\psi+q$}{gq to J/psi+q}}

A representative $t$-channel diagram for the $gq$ subprocess is shown in Fig.~\ref{fig:feynman}(b); the full $gq$ amplitude is not purely $t$-channel, however. The color-octet helicity-resolved amplitudes are tabulated in Eqs.~(A4a)--(A4i) of the same reference and contain, besides $\hat s-M^2$ and $\hat t-M^2$ denominators, the $1/\hat t$ pole of the exchanged gluon
(and, for ${}^3S_1^{[8]}$, a $1/\hat u$ pole). The $c\bar{c}$ helicity components are summed in our evaluation.

\subsubsection{\texorpdfstring{$gg\to J/\psi+g$}{gg to J/psi+g}}

The short-distance squared amplitude for $c\bar{c}[{}^3S_1^{[1]}]$ formation in the $gg$ subprocess is~\cite{Baier:1983va,Gastmans:1987be}
\begin{align}
\label{eq:amp_gg_3S11}
\overline{\sum}|\mathcal{A}(gg\to c\bar{c}[{}^3S_1^{[1]}]\,g)|^2
&= \frac{160\pi^3\alpha_s^3\, M}{81}
\biggl[
  \frac{\hat{s}^2}{[(\hat{t}-M^2)(\hat{u}-M^2)]^2}
+ \frac{\hat{t}^2}{[(\hat{u}-M^2)(\hat{s}-M^2)]^2}
\nonumber \\
&
\qquad\qquad\qquad
+ \frac{\hat{u}^2}{[(\hat{s}-M^2)(\hat{t}-M^2)]^2}
\biggr].
\end{align}

For the color-octet states ${}^1S_0^{[8]}$, ${}^3S_1^{[8]}$, and ${}^3P_J^{[8]}$, the helicity-resolved squared amplitudes are given in Eqs.~(A5a)--(A5i) of Ref.~\cite{Cho:1995ce}. These expressions involve channel-dependent denominator structures reflecting the internal heavy-quark propagators [Fig.~\ref{fig:feynman}(c)]. The $c\bar{c}$ helicity components ($h = 0, 1, 2$) are summed in our evaluation to obtain the unpolarized squared amplitudes.

\subsection{Numerical inputs}
\label{subsec:numerical_inputs}
We use the NNPDF4.0 NNLO PDF set~\cite{NNPDF:2021njg} with central factorization and renormalization scales set to $\mu_F = \mu_R = E_T = \sqrt{M^2+p_T^2}$, following Ref.~\cite{Chao:2012iv}. The $J/\psi$ mass is $M = 2m_c = 3.0\,\mathrm{GeV}$ (with $m_c = 1.5\,\mathrm{GeV}$, as in Refs.~\cite{Ma:2010jj,Chao:2012iv}), used consistently throughout all kinematic expressions. We use $M = 2m_c$ rather than the measured $J/\psi$ mass $3.097\,\mathrm{GeV}$~\cite{ParticleDataGroup:2024cfk} because at leading order in $v$ the $c$ and $\bar{c}$ are produced at rest relative to each other, so that $M = 2m_c$ is what enters the short-distance amplitudes consistently, and because it is the convention in which the LDMEs of Table~\ref{tab:ldme} were extracted~\cite{Chao:2012iv}; the $3\%$ difference is well within the $\mathcal{O}(v^2)$ relativistic corrections neglected at this order. The running coupling $\alpha_s(\mu_R)$ entering the LO matrix elements is taken directly from the NNPDF4.0 NNLO set, which uses three-loop evolution with $\alpha_s(M_Z) = 0.118$, ensuring that the strong coupling in the matrix elements and in the DGLAP evolution of the PDFs is treated consistently. The NRQCD factorization scale is fixed at $\mu_\Lambda = m_c$, consistent with the scheme in which the LDMEs of Table~\ref{tab:ldme} are extracted~\cite{Chao:2012iv}.

Theoretical scale uncertainty is estimated via the standard 7-point variation: $\mu_F, \mu_R \in [E_T/2,\, 2\,E_T]$ with the constraint $1/2 \leq \mu_F/\mu_R \leq 2$, excluding the extreme combinations $(\mu_F, \mu_R) = (E_T/2,\, 2 E_T)$ and $(2 E_T,\, E_T/2)$ which introduce large unphysical logarithms. At leading order in $\alpha_s$, the $\mu_R$ dependence enters through the running coupling $\alpha_s(\mu_R)$, which appears at cubic order in the partonic amplitudes, while the $\mu_F$ dependence enters through the DGLAP-evolved parton distributions $f_{a/p}(x, \mu_F)$; both are uncompensated at this order and the resulting scale variation is expected to be sizable.
For $p_T\lesssim1.4$~GeV the lowest factorization scale, $E_T/2$, lies below the NNPDF4.0 grid ($Q_{\min}=1.65$~GeV), where we use
the LHAPDF extrapolation~\cite{Buckley:2014ana}.

The ingredients of the calculation enter at different perturbative orders: the short-distance matrix elements at leading order, $\mathcal{O}(\alpha_s^3)$; the parton densities from the NNLO NNPDF4.0 set, with NNLO DGLAP evolution and the three-loop $\alpha_s$ that accompanies it; and the LDMEs of Table~\ref{tab:ldme} from an NLO fit.  
Combining LO matrix elements with LDMEs from NLO fits is also the choice of the leading-order analysis to which we compare, which uses the NLO sets of Ref.~\cite{Chao:2012iv} among others~\cite{Chen:2024nfl}, and we follow it,\footnote{Leading-order LDME fits exist~\cite{Cho:1995ce,Sharma:2012dy}, but at LO the ${}^1S_0^{[8]}$ and ${}^3P_J^{[8]}$ short-distance coefficients have nearly identical $p_T$ shapes, so such fits constrain only a linear combination of the two and cannot provide the explicit bracketing of the octet apportionment---the ${}^1S_0^{[8]}$-min and -max sets of Table~\ref{tab:ldme}---on which the sensitivity study of Sec.~\ref{sec:ldme} rests.} 
noting that the LDMEs are effective parameters whose values depend on the order of the fit. Modern LO PDF fits are less reliable than NNLO sets, and at LO the $\mu_F$ dependence is uncompensated whatever the PDF order; we therefore present the scale-variation envelope as an estimate of the perturbative uncertainty rather than a precise bound; an NLO calculation of the correlator, which would restore order matching between the matrix elements and the LDME fit, is deferred to future work.  
The PDF uncertainty, and the treatment of $\mu_F$ below the grid, affect the results less than the scale variation and are not
included in the bands.\footnote{Freezing $\mu_F$ at $Q_{\min}$, or using CT18NNLO~\cite{Hou:2019efy}, whose grid extends below $E_T/2$, gives similar scale bands. PDF variations (the
CT18NNLO Hessian set, or CT18NNLO in place of NNPDF4.0) move the normalized shape by less than the scale variation, and the ratio of
two LDME sets at a common PDF member (Fig.~\ref{fig:ldme_ratio_qec}) by about 1\% or less.}

\begin{table}
\centering
\begin{tabular}{l c c c c}
\hline\hline
Scenario
  & $\LDME{{}^3S_1^{[1]}}$
  & $\LDME{{}^1S_0^{[8]}}$
  & $\LDME{{}^3S_1^{[8]}}$
  & $\LDME{{}^3P_0^{[8]}}/m_c^2$ \\
  & $[\mathrm{GeV}^3]$
  & $[10^{-2}\,\mathrm{GeV}^3]$
  & $[10^{-2}\,\mathrm{GeV}^3]$
  & $[10^{-2}\,\mathrm{GeV}^3]$ \\
\hline
Default              & $1.16$ & $8.9$ & $0.30$ & $0.56$ \\
${}^1S_0^{[8]}$-min  & $1.16$ & $0$   & $1.4$  & $2.4$  \\
${}^1S_0^{[8]}$-max  & $1.16$ & $11$  & $0$    & $0$    \\
\hline\hline
\end{tabular}
\caption{LDME values of Ref.~\cite{Chao:2012iv}, from an NLO fit to Tevatron CDF $J/\psi$ cross section and polarization data: Default is the best fit, ${}^1S_0^{[8]}$-min and ${}^1S_0^{[8]}$-max the two extremes along the direction the data leave undetermined.  The ${}^3P_J^{[8]}$ LDMEs for $J=1,2$ follow by heavy-quark spin symmetry (Sec.~\ref{sec:nrqcd}).}
\label{tab:ldme}
\end{table}

For the LDMEs we adopt the three sets of Chao~et al.~\cite{Chao:2012iv}, obtained from a simultaneous NLO fit to Tevatron CDF $J/\psi$ cross section and polarization data at $p_T>7\,\mathrm{GeV}$, whose extremes switch off in turn $\LDME{{}^1S_0^{[8]}}$ and the spin-triplet octets along the direction that these data leave undetermined:
\begin{itemize}
\item \textbf{Default}: the best-fit central values of Ref.~\cite{Chao:2012iv} with all three color-octet LDMEs nonzero.
\item \textbf{${}^1S_0^{[8]}$-min}: $\LDME{{}^1S_0^{[8]}}=0$, with the cross section data accommodated entirely by the ${}^3S_1^{[8]}$ and ${}^3P_J^{[8]}$ contributions.
\item \textbf{${}^1S_0^{[8]}$-max}: $\LDME{{}^3S_1^{[8]}}=\LDME{{}^3P_J^{[8]}}=0$, with the cross section data accommodated entirely by the ${}^1S_0^{[8]}$ contribution.
\end{itemize}
The color-singlet LDME is fixed at $\LDME{{}^3S_1^{[1]}} = 1.16~\mathrm{GeV}^3$, computed from the $J/\psi$ wave function at the origin
of the Buchm\"uller--Tye potential model, $|R(0)|^2=0.810~\mathrm{GeV}^3$~\cite{Eichten:1995ch,Brambilla:2010cs}, with the NRQCD
normalization $\LDME{{}^3S_1^{[1]}}=\frac{3N_c}{2\pi}|R(0)|^2$~\cite{Bodwin:1994jh}, as in Ref.~\cite{Chao:2012iv}; it takes the same value in
all three sets. The LDME values are collected in Table~\ref{tab:ldme}.

Other NLO extractions realize the two extremes approximately: the fragmentation-improved fits of Bodwin~et al.\ return
$\LDME{{}^1S_0^{[8]}}$ close to the Default value when the prompt data are fitted directly~\cite{Bodwin:2014gia} and close to the
${}^1S_0^{[8]}$-max value when the feed-down is computed separately~\cite{Bodwin:2015iua}, while the pNRQCD determination of
Brambilla~et al.~\cite{Brambilla:2022ayc} lies close to the ${}^1S_0^{[8]}$-min values.\footnote{The global fit of Butenschoen and
Kniehl~\cite{Butenschoen:2011yh}, which extends to $p_T>3\,\mathrm{GeV}$ in hadroproduction and includes photoproduction, sits between
them in $\LDME{{}^1S_0^{[8]}}$ but with a negative $\LDME{{}^3P_0^{[8]}}$. Negative values of this kind are needed to describe the
low-$p_T$ hadroproduction and the photoproduction data, but they make the predicted $J/\psi+\gamma$ cross section negative at large
$p_T$~\cite{Lansberg:2019adr}; an individual channel, on the other hand, depends on the renormalization scheme and scale and is not
itself a physical rate~\cite{Chao:2012iv}. The later refit of Ref.~\cite{Shao:2014yta}, which includes the feed-down explicitly,
scans $\LDME{{}^1S_0^{[8]}}$ between zero and its maximal value in the same way, with non-negative LDMEs.}
We use the Chao~et al.\ sets because their extremes bracket the apportionment of the cross section among the octet
channels explicitly, and it is the sensitivity to that apportionment which we set out to probe.
Heavy-quark spin symmetry relates
$\langle \mathcal{O}^{J/\psi}({}^1S_0^{[8]})\rangle$ to
$\langle \mathcal{O}^{\eta_c}({}^3S_1^{[8]})\rangle$, and the LHCb
measurement of prompt $\eta_c$ production then implies
$\langle \mathcal{O}^{J/\psi}({}^1S_0^{[8]})\rangle \lesssim
1.5\times 10^{-2}~\mathrm{GeV}^3$~\cite{Han:2014jya}, below the Default
and ${}^1S_0^{[8]}$-max values of Table~\ref{tab:ldme}; we nevertheless
retain these sets, since they serve here only as a bracket.

Throughout Secs.~\ref{sec:amplitudes}--\ref{sec:interpretation} we present channel-resolved results using the Default LDMEs, so that each NRQCD channel contributes with a nonzero weight and the full channel-dependent angular-momentum structure can be resolved. The three sets are compared in Sec.~\ref{sec:ldme}, where the differences among them quantify the LDME sensitivity of the QEC measurement.

\subsection{Validation against \textsc{HELAC-onia}}
\label{subsec:helac}

\begin{figure}
\centering
\includegraphics[width=\linewidth]{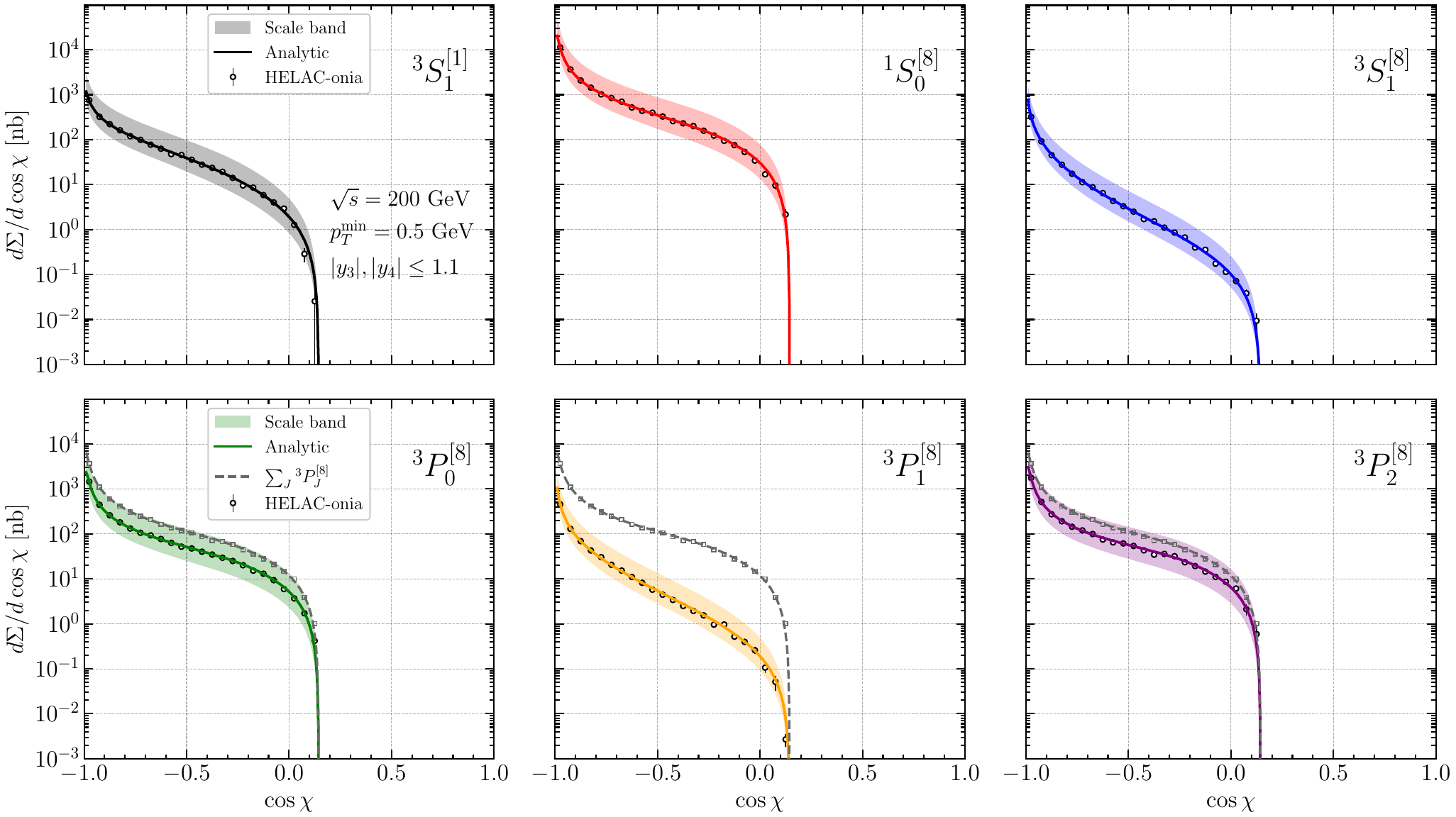}
\caption{Analytic QEC (colored lines, with the 7-point scale band) and \textsc{HELAC-onia} (open markers, statistical
  errors) for the six NRQCD channels at $\sqrt{s}=200\,\mathrm{GeV}$ in the central acceptance, each with its Default LDME
  (Table~\ref{tab:ldme}) and a common $y$-axis.  \textsc{HELAC-onia} is binned in steps of 0.05 and placed at the bin center
  with the analytic shape.  In the ${}^3P_J^{[8]}$ panels the gray dashed curve and open squares show the spin-symmetric sum
  $\sum_J{}^3P_J^{[8]}$ (Sec.~\ref{sec:nrqcd}).}
\label{fig:helac_comparison}
\end{figure}

\textsc{HELAC-onia}~\cite{Shao:2012iz,Shao:2015vga} is a tree-level Monte Carlo generator that implements the full NRQCD partonic matrix elements. 
For each channel we generate parton-level events with the NNPDF4.0 PDF set, the same central scales $\mu_F=\mu_R=E_T$, and the same LDMEs, apply the kinematic cuts of Sec.~\ref{subsec:analytic_QEC}, and compute $\cos\chi$ event-by-event using the Lorentz transformation of Sec.~\ref{sec:kinematics}.
The samples contain between $\sim 5\times 10^4$ and $\sim 3\times10^6$ events, depending on channel and acceptance, the marker scatter in Figs.~\ref{fig:helac_comparison}, \ref{fig:channels_atlascms} and \ref{fig:channels_lhcb} being statistical.
With identical inputs, the comparison tests our implementation---the squared amplitudes, the $\cos\chi$ map with its root finding and Jacobian, and the acceptance integration---but not the common inputs themselves (parton distributions, scales, and LDMEs).

Figure~\ref{fig:helac_comparison} shows the comparison for all NRQCD channels at $\sqrt{s}=200\,\mathrm{GeV}$, presented as a $2 \times 3$ panel grid (one channel per panel) with a common $y$-axis range to facilitate direct comparison of channel magnitudes.
The analytic curves and the bin-center estimates from \textsc{HELAC-onia} agree in both shape and absolute normalization across all six channels, including the mid-backward shoulder, $\cos\chi \in (-0.6, -0.2)$, visible in ${}^1S_0^{[8]}$, ${}^3P_{0,2}^{[8]}$ and (more subtly) ${}^3S_1^{[1]}$. Reproducing the QEC correctly requires summing over the multiple $p_T$ roots of the non-monotonic $p_T \mapsto \cos\chi$ map (Sec.~\ref{sec:kinematics}), a structure absent in the fragmentation approximation. 
The map folds only when the recoil parton lies farther from mid-rapidity than the $J/\psi$ and on the opposite side, which confines the multiple roots to $\cos\chi\lesssim-0.77$ at sPHENIX kinematics ($\lesssim-0.46$ at ATLAS/CMS; see Fig.~\ref{fig:coschi_pt}); in a forward acceptance such as LHCb's, where $y_3$ and $y_4$ are both positive, the condition is never met and the map is monotonic throughout. Retaining only the large-$p_T$ root---the branch that survives in the fragmentation limit---would lower the LDME-weighted total by up to ${\sim}40\%$ as $\cos\chi\to-1$ (up to ${\sim}50\%$ for individual channels) and the integrated $\Sigma$ by ${\sim}12\%$ at sPHENIX (${\sim}30\%$ at ATLAS/CMS, where most of $\Sigma$ lies near the endpoint), while leaving the distribution unchanged above the fold region.

The shoulder lies above $\cos\chi=-0.77$, where the map is single-valued, so it is a feature of the squared amplitudes, not of the folding of the $p_T\mapsto\cos\chi$ map.
The channel-dependent forward/backward population pattern apparent in
Fig.~\ref{fig:helac_comparison}---in particular the contrast between ${}^3S_1^{[8]}$ and ${}^1S_0^{[8]}$, and the differences among the ${}^3P_J^{[8]}$ substates---is interpreted in Sec.~\ref{sec:interpretation} as a consequence of the angular-momentum structure of the production amplitudes.
The spin-symmetric sum $\sum_J{}^3P_J^{[8]}$, overlaid in the ${}^3P_J^{[8]}$ panels, 
is carried in comparable parts by ${}^3P_2^{[8]}$ and ${}^3P_0^{[8]}$ and inherits their common $J\neq1$ shape; the ${}^3P_1^{[8]}$ contribution to the sum is small despite its $3/9$ weight: through the dynamical suppression of Sec.~\ref{subsec:total_prediction} it carries less of the total than ${}^3P_0^{[8]}$, whose matrix element is three times smaller. 

A characteristic feature of Fig.~\ref{fig:helac_comparison} is that
the QEC vanishes well before $\cos\chi = +1$, at $\cos\chi \approx 0.14$
for all six channels, a kinematic consequence of the fiducial acceptance (Sec.~\ref{subsec:wide_rapidity}).
The small-$p_T$ limit of Eq.~\eqref{eq:coschi} reads
\begin{equation}
\label{eq:coschi-pT0-limit}
\lim_{p_T \to 0}\cos\chi(y_3, y_4, p_T)
  \;=\; -\sgn(y_3)\,\tanh(y_3 - y_4),
\end{equation}
so that $\cos\chi \to +1$ requires same-sign $y_3, y_4$ with arbitrarily
large $|\Delta_y|$, which $|y_3|, |\eta_4| \leq 1.1$ bounds; with
$p_T \geq p_T^{\min}$ the maximum is $\cos\chi_{\max}^{\rm fid} \approx 0.14$,
attained at $(y_3, y_4, p_T) \approx (0.44, 1.1, 0.5\,\mathrm{GeV})$
(Appendix~\ref{app:coschi_max}). This unfortunately prevents sPHENIX from accessing the more-forward region where the QEC shape exhibits greater differences between different channels, as we will see later for LHC acceptances (Sec.~\ref{subsec:wide_rapidity}).

\subsection{Relation between the QEC and QpTC}
\label{subsec:qptc}

The kinematic relation between the two observables is transparent at
the LO single-recoil level.  Since the recoil's lab-frame transverse
momentum equals $p_T$ by transverse-momentum conservation, while
its rest-frame energy $E_4^{\rm rest}$ is given by Eq.~\eqref{eq:p4_rest}, the
per-event ratio of QpTC to QEC weights is
\begin{equation}
\label{eq:qptc_qec_ratio_LO}
\frac{p_T/M}{E_4^{\rm rest}/M} \;=\; \frac{M}{p_T + E_T\cosh\Delta_y} \;\leq\; \frac{M}{p_T + E_T},
\end{equation}
with the upper bound reached only at $\Delta_y = 0$. The QpTC weight is strictly
smaller than the QEC weight for any $p_T > 0$, with the suppression
amplified at $|\Delta_y| > 0$ through the $\cosh\Delta_y$ factor.

\begin{figure}
\centering
\includegraphics[width=\linewidth]{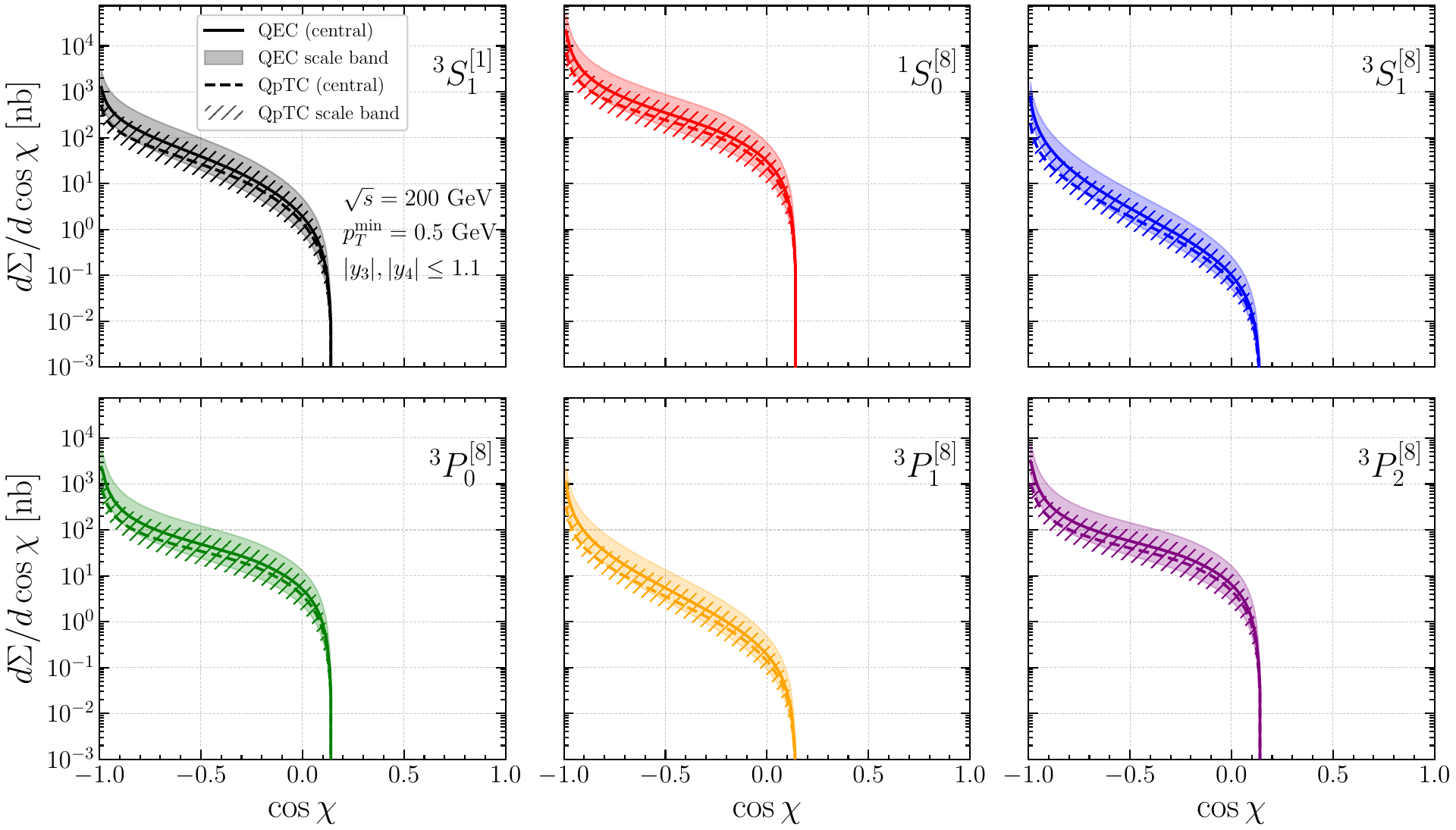}
\caption{Analytic QEC (solid, filled bands) and the QpTC of Ref.~\cite{Zhang:2026tjo} (dashed, hatched bands) for the six channels at $\sqrt{s}=200\,\mathrm{GeV}$ in the central acceptance; bands: 7-point scale variation.  Their ratio is set by Eq.~\eqref{eq:qptc_qec_ratio_LO} (Sec.~\ref{subsec:qptc}).}
\label{fig:qec_vs_qptc}
\end{figure}

Figure~\ref{fig:qec_vs_qptc} displays the analytic QEC and QpTC
predictions side-by-side for all six NRQCD channels at
$\sqrt{s}=200\,\mathrm{GeV}$ with the fiducial cuts of
Sec.~\ref{subsec:analytic_QEC}.  In this acceptance the QpTC reproduces the QEC shape---both are
backward-peaked---but with a $\cos\chi$-dependent suppression set by the
per-event weight ratio [Eq.~\eqref{eq:qptc_qec_ratio_LO}]: strongest at
the backward peak, where $\cos\chi\to-1$ is reached at large $p_T$ and the
ratio falls to $\sim 0.3$--$0.4$, and weakening toward mid-$\cos\chi$ to a
broad maximum of $\sim 0.75$ before easing to $\sim 0.7$ at the forward
cutoff.
The two weightings probe formally distinct quantities---by the LO identity
Eq.~\eqref{eq:E4pp_shat} the QEC weight measures the partonic center-of-mass
energy excess $(\hat{s}-M^2)/2M^2$, while the QpTC weight measures only its
transverse component---but the steeply falling $p_T$ spectrum and the fiducial
cap on $|\Delta_y|$ lock the two together in the central acceptance, so the
suppression is a smooth kinematic factor and the QpTC carries the same
channel information as the QEC at this order.

\subsection{Total LDME-weighted predictions at sPHENIX}
\label{subsec:total_prediction}

\begin{figure}
\centering
\includegraphics[width=\linewidth]{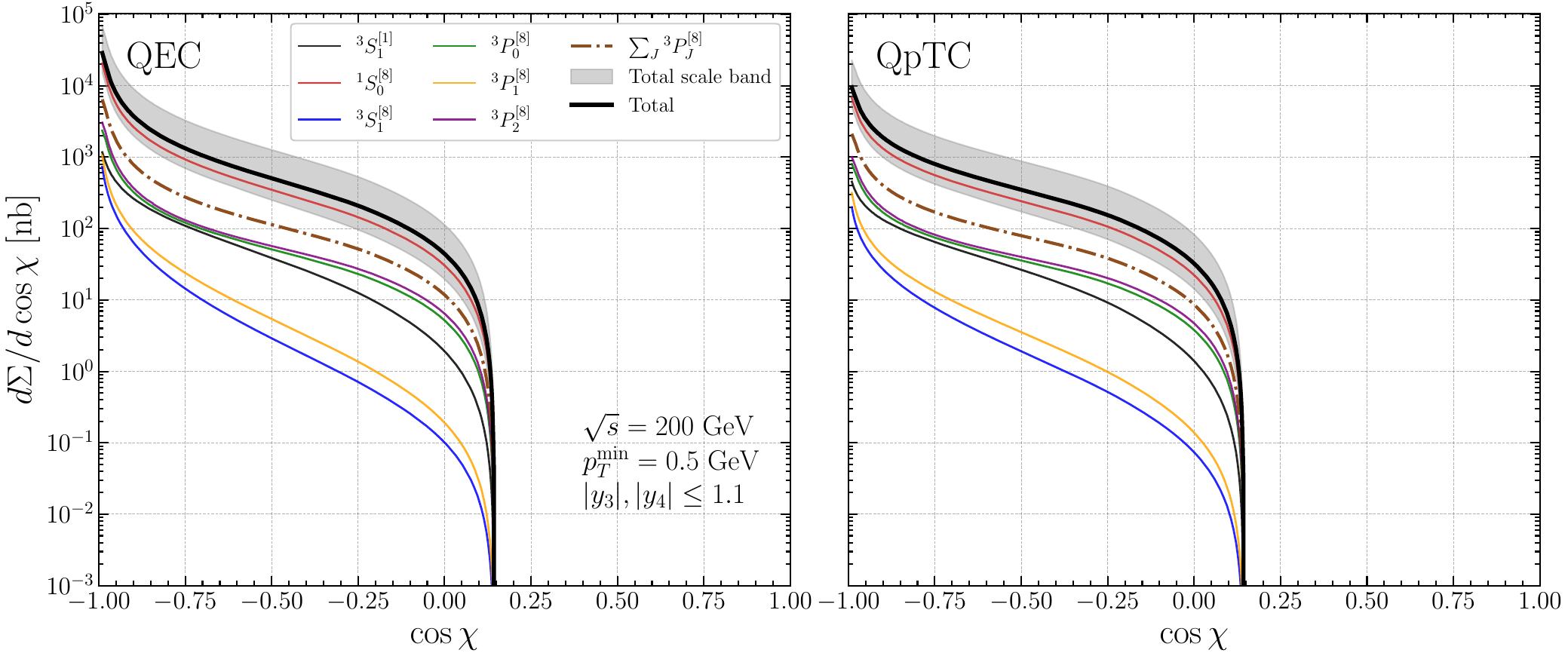}
\caption{LDME-weighted total QEC (left) and QpTC (right) at $\sqrt{s}=200\,\mathrm{GeV}$ with the Default LDMEs: the channel
  sum (thick) with its 7-point scale band, the six channel contributions, and the spin-symmetric sum
  $\sum_J{}^3P_J^{[8]}$ (dash-dotted).  The total is dominated by ${}^1S_0^{[8]}$.}
\label{fig:total_qec_qptc}
\end{figure}

Summing the channel-resolved analytic baselines of
Secs.~\ref{subsec:helac}--\ref{subsec:qptc} with the Default LDMEs of
Table~\ref{tab:ldme} yields the inclusive prediction
\begin{equation}
\label{eq:total_qec_qptc}
\frac{d\Sigma^{\rm tot}}{d\cos\chi}
\;=\;\sum_{n}\,
  \frac{d\Sigma^{[n]}}{d\cos\chi}
\;\bigg|_{\LDME{n}=\,\text{Default}},
\end{equation}
where $d\Sigma^{[n]}/d\cos\chi$ denotes the channel-$n$ contribution to either the QEC or the QpTC, and
$n \in \{{}^3S_1^{[1]}, {}^1S_0^{[8]}, {}^3S_1^{[8]}, {}^3P_J^{[8]}\}$
with $J=0,1,2$.
Figure~\ref{fig:total_qec_qptc} displays the channel decomposition
together with the total for both observables.

Integrated over the full range, $\cos\chi\ge-1$ (the distribution rises as $(1+\cos\chi)^{-1/2}$ toward the endpoint but is integrable), the cross section under the Default LDMEs is $\Sigma_{\rm QEC} \simeq 2.1 \times 10^3$ nb, with the
7-point correlated scale variation spanning a factor of $\sim 0.5$
to $\sim 2.4$ relative to this central value.

Under the Default-LDME assignment, the ${}^1S_0^{[8]}$ channel
contributes ${\sim}70\%$ of the integrated total, followed by
${}^3P_2^{[8]} \,({\sim}10\%)$, ${}^3P_0^{[8]} \,({\sim}9\%)$, and
${}^3S_1^{[1]} \,({\sim}6\%)$.  The two color-octet $J=1$ channels
${}^3S_1^{[8]}$ and ${}^3P_1^{[8]}$ together contribute ${\sim}5\%$.
For ${}^3S_1^{[8]}$ this follows from its small LDME
(Table~\ref{tab:ldme}); for ${}^3P_1^{[8]}$ it does not, since heavy-quark
spin symmetry gives it three times the ${}^3P_0^{[8]}$ matrix element and it
nonetheless falls below the ${\sim}9\%$ of ${}^3P_0^{[8]}$.  That suppression
is dynamical, and is traced to the $J=1$ selection rule in
Sec.~\ref{subsec:jstructure}.
Under the ${}^1S_0^{[8]}$-min and ${}^1S_0^{[8]}$-max sets of Table~\ref{tab:ldme} this apportionment changes completely---the total is carried by the ${}^3P_J^{[8]}$ and ${}^3S_1^{[8]}$ octets in the former and by ${}^1S_0^{[8]}$ alone in the latter---yet, because the dominant channels in all three sets share the $J\neq1$ shape, the summed distribution changes far less than its composition.
The degeneracy of the inclusive $p_T$ fit therefore largely survives in the QEC total within the central acceptance; whether the normalized shape can lift it, here and at the wider acceptances, is the subject of Sec.~\ref{sec:ldme}.

The Default-LDME total inherits the mid-backward shoulder of the dominant
${}^1S_0^{[8]}$ channel, and the six channels sort by how steeply they fall toward
the forward cutoff---organized by the total angular momentum $J$ of the intermediate
$c\bar{c}$ pair, with the color singlet between the two octet groups.  This pattern,
a property of the partonic squared amplitudes (Sec.~\ref{sec:interpretation}), is
compressed within the central acceptance and exposed as the forward acceptance widens
(Sec.~\ref{subsec:wide_rapidity}).

% ===========================================================================
\section{Role of Soft Gluon Radiation}
\label{sec:soft}

The color-octet $c\bar{c}$ pair must emit soft gluons to form the physical $J/\psi$, and a complete characterization of this hadronization step requires the detailed Monte Carlo studies pursued in Ref.~\cite{Zhang:2026tjo}.
Having established the analytic partonic baselines for both the QEC and the QpTC, validated against \textsc{HELAC-onia} in Sec.~\ref{sec:amplitudes}, we estimate the partonic-level impact of soft radiation at sPHENIX kinematics by layering a simple isotropic two-body decay model $c\bar{c}^{[8]}\to J/\psi+g_{\rm soft}$ on top of those baselines.

\begin{figure}
\centering
\includegraphics[width=\linewidth]{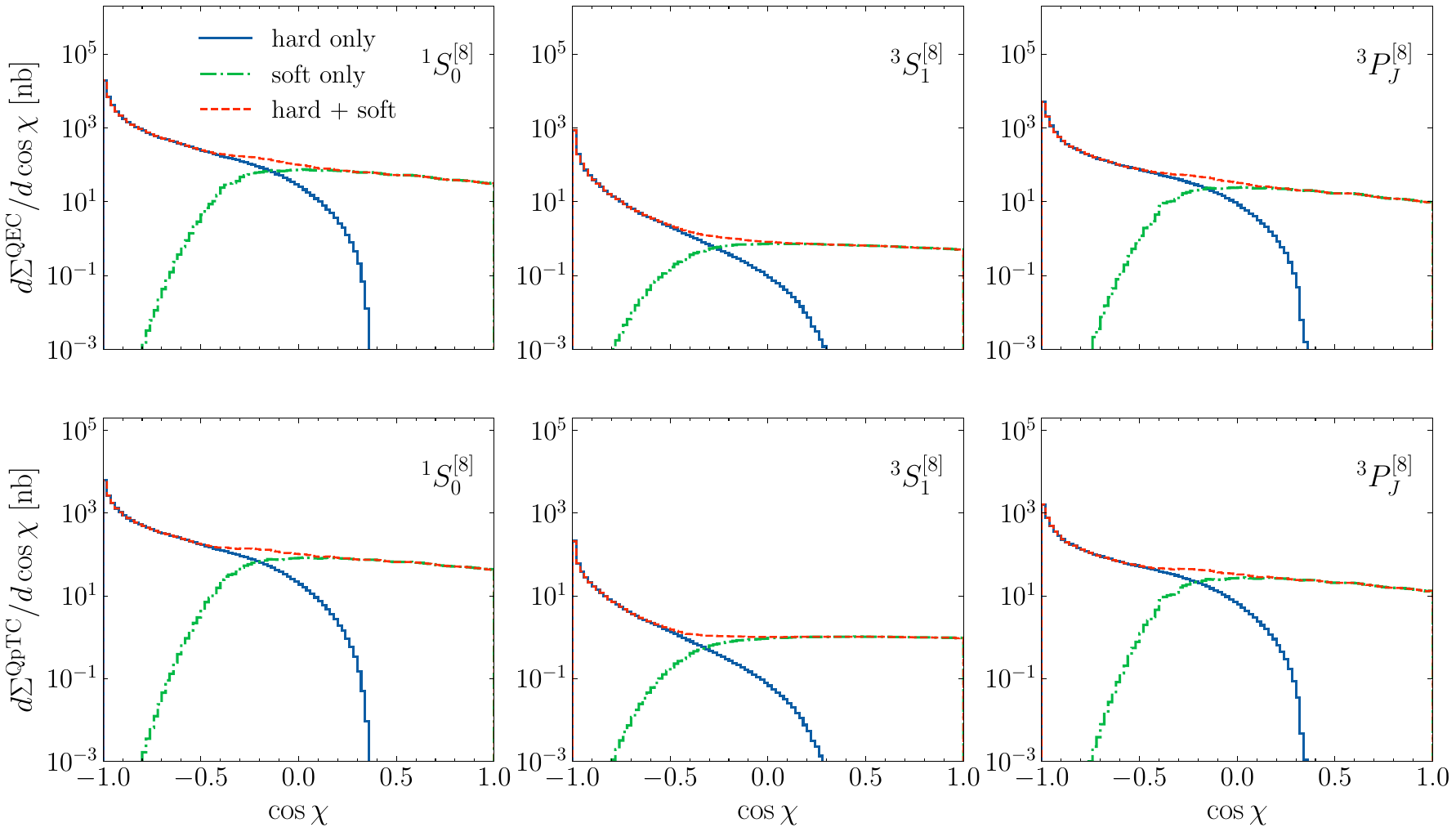}
\caption{QEC (top) and QpTC (bottom) for ${}^1S_0^{[8]}$, ${}^3S_1^{[8]}$ and ${}^3P_J^{[8]}$ at $\sqrt{s}=200\,\mathrm{GeV}$ from \textsc{HELAC-onia} $c\bar{c}^{[8]}$ events ($M_{c\bar{c}}=3.2$~GeV) with the isotropic decay $c\bar{c}^{[8]}\to J/\psi+g_{\rm soft}$: hard recoil, soft gluon, and their sum.  Both particles have $|\eta|\leq1.1$, with $p_T\geq0.5$~GeV (hard) and $0.2$~GeV (soft).  Scale bands are omitted; the hard ones are shown in Fig.~\ref{fig:helac_comparison}, and the soft contribution carries a model uncertainty instead.}
\label{fig:soft_radiation}
\end{figure}

To estimate the non-perturbative soft-gluon contribution from color-octet hadronization at $\sqrt{s}=200\,\mathrm{GeV}$,
we generate $pp\to c\bar{c}^{[8]}+X$ events in \textsc{HELAC-onia}
with the $c\bar{c}^{[8]}$ treated as a particle of mass
$M_{c\bar{c}} = 2 m_c^{[8]} = 3.2\,\mathrm{GeV}$, slightly above
$M$ so that its decay into $J/\psi$ and a soft gluon is kinematically allowed.  Here $m_c^{[8]} = 1.6\,\mathrm{GeV}$
is adopted only for the octet intermediate state, slightly above the
input $m_c = 1.5\,\mathrm{GeV}$ used elsewhere.\footnote{The mass-splitting parameter $M_{c\bar{c}}-M = 0.2\,\mathrm{GeV}$
adopted here corresponds to the default \textsc{Pythia\,8}
\texttt{Onia:massSplit} setting.  Ref.~\cite{Zhang:2026tjo} reports an enhancement of up to ${\sim}60\%$ of the hadron-level correlator at $\cos\chi>0$ ($\sqrt{s}=500\,\mathrm{GeV}$, $7<p_T<15\,\mathrm{GeV}$) when this splitting is varied between $0.2$ and $0.8\,\mathrm{GeV}$. In our setup a larger splitting increases the soft-gluon energy
$E_g = (M_{c\bar{c}}^2-M^2)/(2M_{c\bar{c}})$ and amplifies the
forward-hemisphere effect identified in Fig.~\ref{fig:soft_radiation}, more strongly in the QpTC, where the soft share is larger.
A quantitative splitting-parameter scan at sPHENIX kinematics is
left to future work.}
We then decay each generated $c\bar{c}^{[8]}$ isotropically into $J/\psi + g_{\rm soft}$ in its rest frame, averaging each event over the decay direction.
Since $M_{c\bar{c}} > M = 3.0\,\mathrm{GeV}$, the decay is kinematically allowed, with
\begin{equation}
\label{eq:soft_decay}
E_{g} = \frac{M_{c\bar{c}}^2 - M^2}{2M_{c\bar{c}}} \approx 0.194\,\mathrm{GeV},
\qquad
E_{J/\psi} = \frac{M_{c\bar{c}}^2 + M^2}{2M_{c\bar{c}}} \approx 3.006\,\mathrm{GeV},
\end{equation}
both evaluated in the $c\bar{c}^{[8]}$ rest frame. The $c\bar{c}^{[8]}$ is then replaced by the resulting $J/\psi + g_{\rm soft}$ pair, and the QEC and QpTC are evaluated in the rest frame of the new $J/\psi$, with both the hard recoil parton and the soft gluon contributing to the energy flow.

Figure~\ref{fig:soft_radiation} displays the hard-only, soft-only, and hard-plus-soft contributions to the QEC and QpTC for the three color-octet channels at $\sqrt{s}=200\,\mathrm{GeV}$.
For both observables the soft-gluon contribution is subdominant over most of the backward hemisphere, down to $\cos\chi\approx-0.15$ to $-0.3$ depending on channel and observable, and takes over toward the fiducial cutoff of the hard contribution; beyond $\cos\chi\approx0.32$ the hard contribution falls below $10^{-4}$ of the soft one.\footnote{The hard boundary moves from $\cos\chi_{\max}^{\rm fid} \approx 0.14$ [Eq.~\eqref{eq:coschi-max-fid}] to $0.36$ in this setup: the $c\bar{c}^{[8]}$ mass $M_{c\bar{c}}=3.2$ GeV alone shifts it to $\sim 0.17$, and the boost from the $c\bar{c}^{[8]}$ to the new $J/\psi$ rest frame, which depends on the decay direction, extends it to $0.36$, reached only in rare decay configurations.}
The soft gluon carries only $(M_{c\bar{c}}^2-M^2)/(2M)\approx0.207\,\mathrm{GeV}$ in the $J/\psi$ rest frame, a QEC weight $E_g^{\rm rest}/M\approx0.069$ against $\mathcal{O}(1$--$10)$ for the backward hard recoil; it dominates the forward region because the hard contribution is kinematically excluded there, not because it is large.
The soft-only distribution is not flat in $\cos\chi$ even though the emission is isotropic: a gluon emitted backward
in the $J/\psi$ rest frame is boosted to low lab-frame momentum and fails the $p_T\geq0.2$ GeV cut, which suppresses
the soft contribution as $\cos\chi\to-1$.
Integrated over the fiducial, the soft contribution amounts to $\sim3$--$9\%$ of the hard QEC and $\sim14$--$23\%$ of the hard QpTC across the three channels; the larger QpTC ratio arises because the hard-recoil QpTC weight is suppressed by $M/[p_T+E_T\cosh\Delta_y]\sim0.3$--$0.4$ at the backward peak [Eq.~\eqref{eq:qptc_qec_ratio_LO}] while the soft-gluon QpTC weight stays comparable to its QEC counterpart.

The model is deliberately schematic.  It treats the octet-to-singlet transition as a single isotropic emission, ignoring its multi-gluon character, its dependence on the $c\bar{c}$ spin and color configuration, and any correlation with the hard-recoil direction.  Because an isotropic emission carries no angular structure of its own, it enters every octet channel with the same $\cos\chi$ shape and, beyond the hard cutoff, 
flattens the channel-to-channel contrast rather than adding to it
(Fig.~\ref{fig:soft_radiation}). That flatness does not make the soft contribution insensitive to the LDMEs, however: its height is fixed by the summed color-octet rate, so the forward region carries the normalization of the octet component even where it no longer resolves the individual channel shapes.  Combining the two---the hard channel shapes below the kinematic cutoff with the soft contribution above it---is what the forward window offers, and is the sense in which it probes the color-octet hadronization studied in Refs.~\cite{Chen:2024nfl,Zhang:2026tjo}.

In the central acceptance, the forward region $\cos\chi\in(0.32,1)$ offers direct access to the non-perturbative soft emission, essentially free of the hard $2\to2$ contribution---a probe of the octet hadronization~\cite{Chen:2024nfl}, once the higher-order real emission that also reaches it is accounted for.

The analysis of Ref.~\cite{Chen:2024nfl}, however, works in a fragmentation picture---accounting for single hard-gluon, single or double soft-gluon emissions---but omits the recoil-parton contribution to the energy flow, deferring it to future work; our full cross-section-level calculation retains it, and shows that with a suitably chosen acceptance for the $J/\psi$ and recoil parton the hard contribution itself reaches substantially into the forward region, where the QEC and QpTC shapes carry the total-angular-momentum structure of the intermediate $c\bar{c}$ state---and, through the channel mix, part of the sensitivity to the LDME sets of Table~\ref{tab:ldme}.
The following sections develop this perturbative hard-radiation sensitivity.

% ===========================================================================
\section{Angular-Momentum Hierarchy and its Kinematic Dependence}
\label{sec:interpretation}

The backward hemisphere of the QEC, dominated over most of its range by hard radiation (Sec.~\ref{sec:soft}), carries a signature of the total angular momentum of the intermediate $c\bar{c}$ Fock state.  
We first compare the channel shapes at the rapidity acceptances of the three experiments (Sec.~\ref{subsec:wide_rapidity}), then trace the grouping they reveal to the partonic squared amplitudes (Sec.~\ref{subsec:jstructure}), and finally vary the collinear cutoff $p_T^{\min}$ (Sec.~\ref{subsec:pt_tuning}).

\subsection{Progressively wider acceptances: sPHENIX, ATLAS/CMS, and LHCb}
\label{subsec:wide_rapidity}

\begin{figure}
\centering
\includegraphics[width=\linewidth]{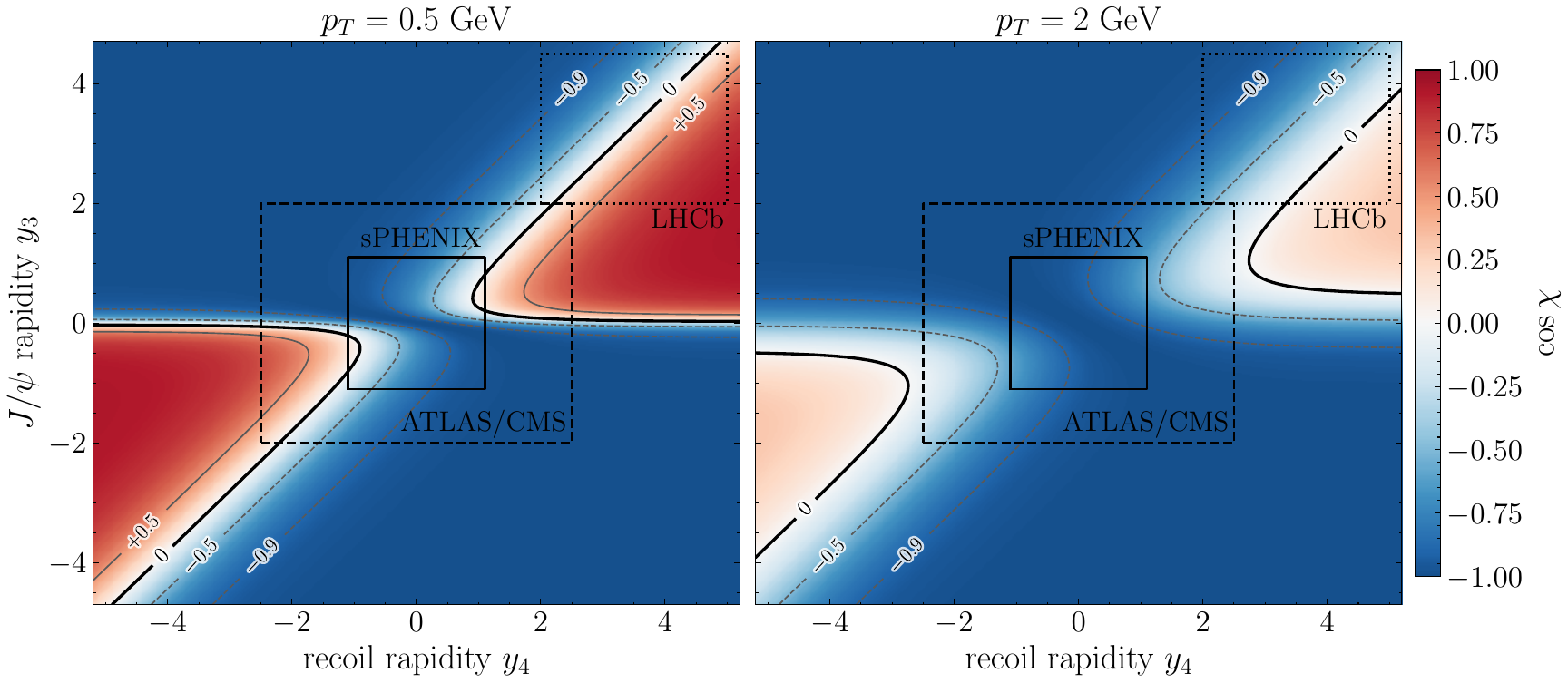}
\caption{$\cos\chi$ of the leading-order recoil, Eq.~\eqref{eq:coschi}, as a function of the recoil and $J/\psi$ rapidities at
  $p_T=0.5$~GeV (left) and $2$~GeV (right).  Thick line: $\cos\chi=0$; thin contours: $\cos\chi=-0.9$, $-0.5$, $+0.5$.  Boxes:
  the fiducial acceptances of sPHENIX ($|y_3|,|y_4|\leq1.1$), ATLAS/CMS ($|y_3|\leq2.0$, $|\eta_4|\leq2.5$), and LHCb
  ($2\leq y_3\leq4.5$, $2\leq\eta_4\leq5$).}
\label{fig:coschi_map}
\end{figure}

In the central acceptance the six channel shapes of Fig.~\ref{fig:helac_comparison} are compressed below $\cos\chi\approx0.14$ (Sec.~\ref{subsec:helac}).  Which acceptances reach farther into the forward hemisphere follows from a simple property of the boost.  A massless particle at laboratory angle $\vartheta$ to the $J/\psi$ flight direction has $\cos\chi=(\cos\vartheta-\beta)/(1-\beta\cos\vartheta)$ in the $J/\psi$ rest frame, with $\beta$ the $J/\psi$ velocity, so $\cos\chi>0$ is the laboratory cone $\sin\vartheta<M/E$ around the $J/\psi$ ($E$ its laboratory energy)---the dead cone of the heavy quarks~\cite{Dokshitzer:1991fd,Chen:2024nfl}.
The recoil parton is not radiated by the heavy quarks and can enter this cone, but since it balances the $J/\psi$ in transverse momentum it does so only when both travel close to the same beam, with same-sign rapidities and $|y_4|>|y_3|$; as $p_T$ grows the cone closes and the recoil is confined to $\cos\chi\approx-1$.
Figure~\ref{fig:coschi_map} shows the resulting forward region, a same-side wedge in the rapidity plane that shrinks with $p_T$: the central acceptance only touches its edge, while the ATLAS/CMS and LHCb acceptances reach progressively deeper into it.

The wider acceptances are evaluated at $13\,\mathrm{TeV}$, so we separate the effect of $\sqrt{s}$ first.  At fixed acceptance
$\cos\chi$ and the partonic cross section depend only on $(p_T,y_3,y_4)$ [Eqs.~\eqref{eq:mandelstam} and \eqref{eq:coschi}], so
the collision energy enters only through the parton luminosity.  
Raising $\sqrt{s}$ increases that luminosity most at large $\hat s=x_1x_2s$, which only the backward region reaches 
[forward $\cos\chi$ requires small $p_T$, Fig.~\ref{fig:coschi_map}, and $\hat s-M^2=2p_T(p_T+E_T\cosh\Delta_y)$ vanishes with $p_T$, Eq.~\eqref{eq:mandelstam}], and therefore tilts every channel toward
$\cos\chi=-1$. 
The channel shapes, set by the partonic amplitudes (Sec.~\ref{subsec:jstructure}), survive this reweighting,
and the partonic mix behind them is stable: forward of $\cos\chi\approx-0.3$, the region that carries the channel-discriminating
information, the $gq/gg$ ratio agrees between $200\,\mathrm{GeV}$ and $13\,\mathrm{TeV}$ to about $10\%$.

The central sPHENIX acceptance $|y_3|,|\eta_4|\leq 1.1$ used so far
confines $|\Delta_y|\leq 2.2$ and caps the forward reach at
$\cos\chi_{\max}^{\rm fid}\approx 0.14$
[Eq.~\eqref{eq:coschi-max-fid}].  
The two wider LHC acceptances extend this reach substantially.  At
$p_T^{\min}=0.5\,\mathrm{GeV}$, the more central ATLAS/CMS acceptance
($|y_3|\leq 2.0$, $|\eta_4|\leq 2.5$)\footnote{We take $|y_3|\leq2.0$ from the ATLAS $J/\psi$ measurement at
$13\,\mathrm{TeV}$~\cite{ATLAS:2023qnh} and $|\eta_4|\leq2.5$, the tracker coverage of both experiments.
The measured cross sections start at $p_T=8\,\mathrm{GeV}$ at ATLAS~\cite{ATLAS:2023qnh} and at
$20\,\mathrm{GeV}$, within $|y|<1.2$, at CMS~\cite{CMS:2017dju}, so with $p_T^{\min}=0.5\,\mathrm{GeV}$ this
acceptance is representative of the detectors rather than of an existing measurement; the LHCb measurement
covers $p_T<14\,\mathrm{GeV}$ in $2<y<4.5$~\cite{LHCb:2015foc}, as our forward acceptance does.} reaches
$\cos\chi_{\max}^{\rm fid}\approx 0.73$, while the forward LHCb window
($2\leq y_3\leq 4.5$, $2\leq \eta_4\leq 5$) \cite{LHCb:2015foc} reaches
$\cos\chi_{\max}^{\rm fid}\approx 0.91$ (Appendix~\ref{app:coschi_max}).
The three fiducial acceptances thus form a sequence of increasing forward reach, $0.14\to 0.73\to 0.91$, that progressively 
opens the forward $\cos\chi$ region in which the channel shapes differ most.
 
\begin{figure}
\centering
\includegraphics[width=\linewidth]{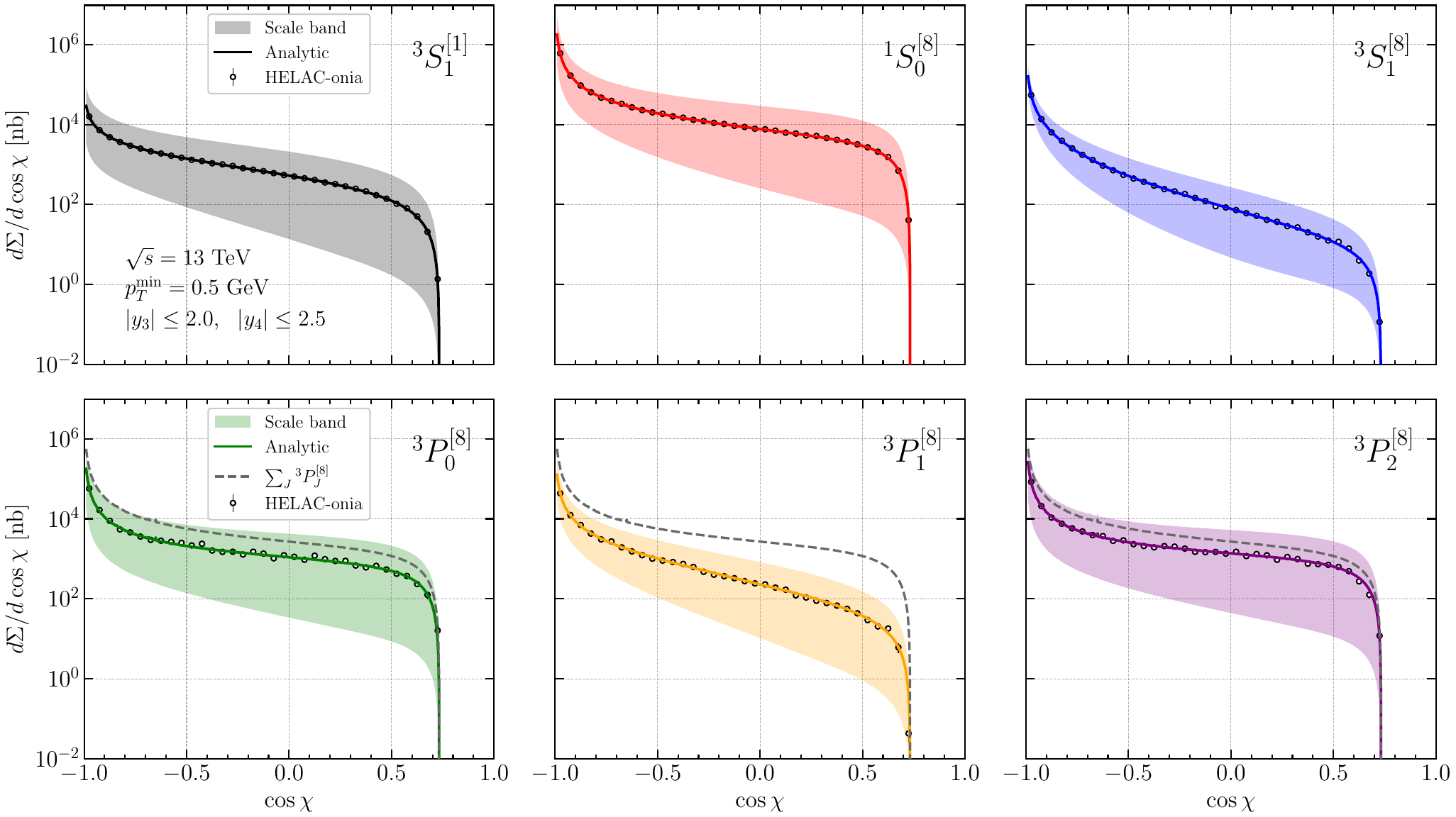}
\caption{As Fig.~\ref{fig:helac_comparison}, at the ATLAS/CMS acceptance ($\sqrt{s}=13\,\mathrm{TeV}$, $|y_3|\leq2.0$,
  $|\eta_4|\leq2.5$, $p_T^{\min}=0.5\,\mathrm{GeV}$), which reaches $\cos\chi_{\max}^{\rm fid}\approx0.73$; the
  $\sum_J{}^3P_J^{[8]}$ sum is analytic only.  The \textsc{HELAC-onia} statistical errors reflect the Monte Carlo sample sizes.}
\label{fig:channels_atlascms}
\end{figure}
 
\begin{figure}
\centering
\includegraphics[width=\linewidth]{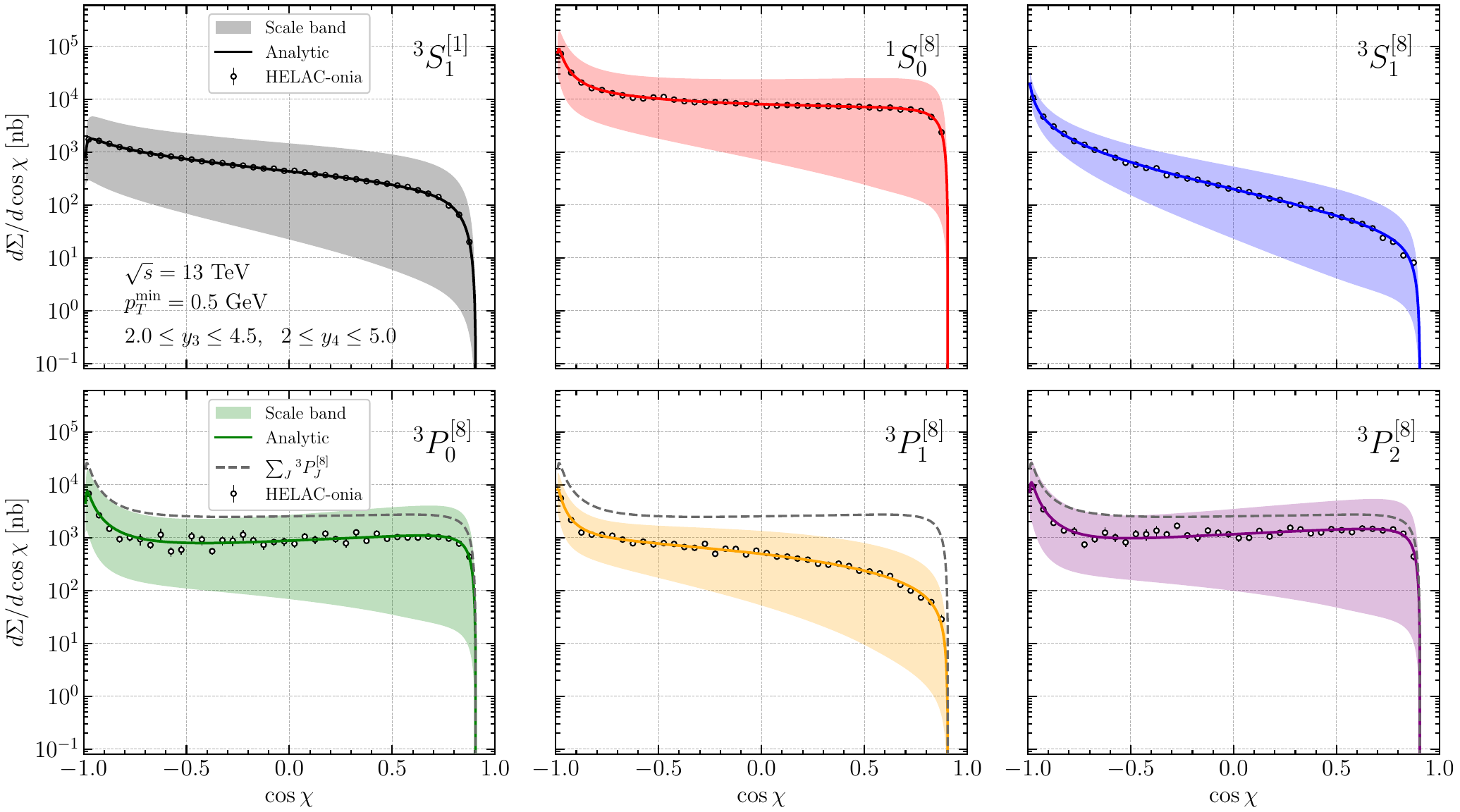}
\caption{As Fig.~\ref{fig:channels_atlascms}, at the forward LHCb acceptance ($\sqrt{s}=13\,\mathrm{TeV}$, $2\leq y_3\leq4.5$,
  $2\leq\eta_4\leq5$)~\cite{LHCb:2015foc}, which reaches $\cos\chi_{\max}^{\rm fid}\approx0.91$.}
\label{fig:channels_lhcb}
\end{figure}
 
Figures~\ref{fig:channels_atlascms} and~\ref{fig:channels_lhcb} show the
six channel-resolved QEC shapes at the ATLAS/CMS and LHCb acceptances;
the sPHENIX counterpart is Fig.~\ref{fig:helac_comparison}.  The channel
discrimination improves markedly as the forward window opens.  The $J=1$ octets
(${}^3S_1^{[8]}$, ${}^3P_1^{[8]}$) and the $J\neq1$ channels
(${}^1S_0^{[8]}$, ${}^3P_{0,2}^{[8]}$), compressed under the central cut, separate visibly once $\cos\chi$ extends past the sPHENIX
cutoff: the $J=1$ octets continue their steep monotonic fall while the $J\neq1$ channels stay flat well into the forward hemisphere. The color-singlet
${}^3S_1^{[1]}$, whose weak central shoulder makes it resemble the $J\neq1$
octets in the narrow window, steepens in the newly opened forward region,
less steeply than ${}^3S_1^{[8]}$ and about as steeply as ${}^3P_1^{[8]}$, consistent with its $J=1$ assignment
(Sec.~\ref{subsec:jstructure}).
The contrast is already clear at ATLAS/CMS and most pronounced at LHCb.
Two remarks are in order.  First, the gain is not in the mid-backward region, where Figs.~\ref{fig:channels_atlascms} and~\ref{fig:channels_lhcb} resemble Fig.~\ref{fig:helac_comparison}, but in the newly opened interval $\cos\chi\gtrsim0.14$, absent at sPHENIX altogether, across which, at LHCb, the $J=1$ octets fall by about an order of magnitude while the $J\neq1$ channels change by less than a factor of 2. Second, the scale bands at $13\,\mathrm{TeV}$ are far wider than at $200\,\mathrm{GeV}$---up to two to three orders of magnitude in the forward region---because the $\mu_F$ variation acts on the steep small-$x$ gluon density and the leading-order cross section carries no compensating collinear counterterm.  They widen toward the forward end and so change the shape as well as the normalization; being largely common to all channels, they cancel to a much larger extent in ratios formed at a common scale (Sec.~\ref{sec:ldme}).

The channels thus fall into two groups, the $J=1$ octets and the $J\neq1$ channels, with the color singlet resembling the $J\neq1$ channels in the central window and the $J=1$ octets in the forward region; the next subsection traces this grouping to the partonic squared amplitudes.

\subsection{Angular-momentum origin of the channel grouping}
\label{subsec:jstructure}

\begin{figure}
\centering
\includegraphics[width=\linewidth]{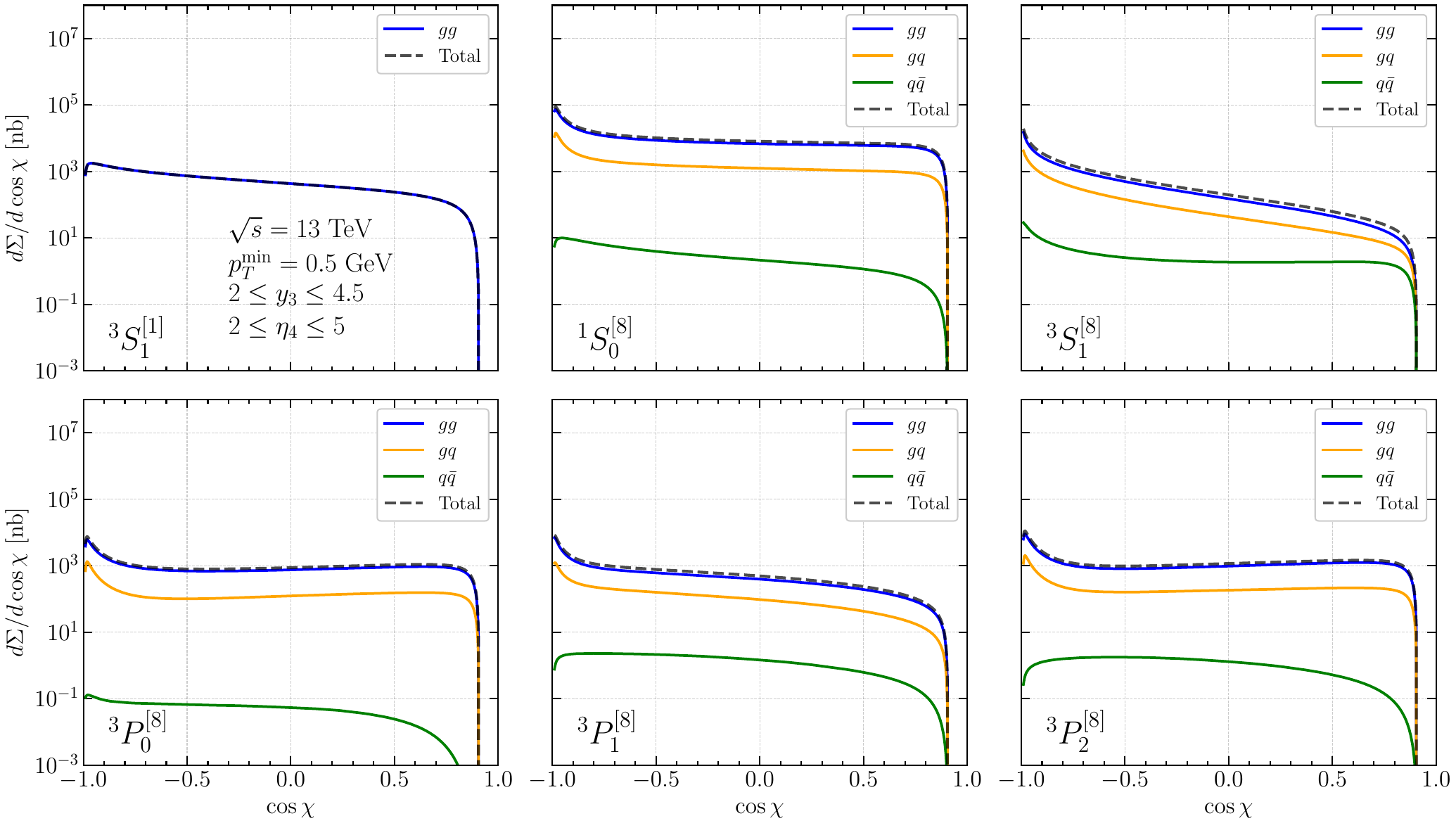}
\caption{QEC decomposed by partonic subprocess for the six channels at the forward LHCb acceptance ($\sqrt{s}=13\,\mathrm{TeV}$, $2\leq y_3\leq4.5$, $2\leq\eta_4\leq5$, $p_T^{\min}=0.5\,\mathrm{GeV}$), with the Default LDMEs, in the layout of Fig.~\ref{fig:helac_comparison}.  The quark-initiated subprocesses vanish for ${}^3S_1^{[1]}$ at this order (Sec.~\ref{sec:amplitudes}).}
\label{fig:channel_breakdown}
\end{figure}

The grouping of Sec.~\ref{subsec:wide_rapidity} is a property of the partonic squared amplitudes, seen most directly in the decomposition of the QEC into the partonic subprocesses $q\bar{q}$, $gq$, and $gg$ for each NRQCD intermediate state (Fig.~\ref{fig:channel_breakdown}).
 
Figure~\ref{fig:channel_breakdown} shows this decomposition at the LHCb acceptance, where the forward range is longest; at sPHENIX the same pattern is compressed below $\cos\chi\approx0.14$.  The $q\bar{q}$ subprocess is everywhere the smallest of the three,
so the overall QEC shapes are determined by the $gg$ and $gq$ subprocesses, in each of which the channels group as follows:
\begin{itemize}
\item Color-octet $J=1$ channels (${}^3S_1^{[8]}$, ${}^3P_1^{[8]}$): a steady
  decrease from $\cos\chi=-1$ to the forward cutoff, in each
  $gg$ and $gq$ subprocess individually; between $\cos\chi=-0.5$ and $0.8$ the channel totals fall by factors of about 35 and 10.
\item Color-octet $J\neq 1$ channels (${}^1S_0^{[8]}$, ${}^3P_0^{[8]}$,
  ${}^3P_2^{[8]}$): a plateau---a range of nearly constant height, in contrast to the steady fall of the $J=1$ octets---from the mid-backward region to the forward cutoff, in each $gg$ and $gq$ subprocess; the totals change by less than a factor of 2 between $\cos\chi=-0.5$ and $0.8$.  In the central acceptance only its onset is visible, as a shoulder across $\cos\chi\in(-0.6,-0.2)$ followed by the drop at the cutoff $\cos\chi\approx0.14$ (Fig.~\ref{fig:helac_comparison}).
\item Color-singlet ${}^3S_1^{[1]}$ ($J=1$): a fall comparable to that of ${}^3P_1^{[8]}$, by a factor of about 9 between $\cos\chi=-0.5$ and $0.8$, while in the narrow central window a weak mid-backward shoulder makes it resemble the $J\neq 1$ octets.
\end{itemize}

This grouping is built into the partonic squared amplitudes rather than emerging
from the convolution with the parton distributions, and its origin is a selection
rule on the initial state, the one that also governs the small-$p_T$ behavior of quarkonium
production~\cite{Petrelli:1997ge,Sun:2012vc}.  
At $\mathcal{O}(\alpha_s^2)$ the $2\to1$ process $gg\to c\bar{c}[n]$ is open for ${}^1S_0^{[8]}$, ${}^3P_0^{[8]}$ and ${}^3P_2^{[8]}$ but closed for every $J=1$ state.  For the color-singlet ${}^3S_1^{[1]}$ and for ${}^3P_1^{[8]}$ this is the Landau--Yang theorem~\cite{Landau:1948kw,Yang:1950rg}, which applies to two on-shell gluons in a color-singlet or color-symmetric state: they cannot form a $J=1$ system.  For the color-octet vector ${}^3S_1^{[8]}$ the theorem does not strictly apply---its color factor, proportional to $f^{abc}$, is antisymmetric under the interchange of the two gluons---but the tree-level amplitude vanishes nonetheless~\cite{Petrelli:1997ge}, while no symmetry forbids the process beyond tree level~\cite{Cacciari:2015ela}.  Since the present calculation is tree level throughout, all three $J=1$ channels are closed to $gg$ here.  Among $2\to1$ processes, the only one open to a $J=1$ state is then $q\bar{q}\to{}^3S_1^{[8]}$; the color-singlet ${}^3S_1^{[1]}$ has no $2\to1$ production process at all, $q\bar{q}\to{}^3S_1^{[1]}$ being forbidden by color and $C$ parity.

\begin{table}
\centering
\setlength{\tabcolsep}{9pt}
\begin{tabular}{lccccc}
\hline\hline
        &     & \multicolumn{3}{c}{$p_T\to0$} & $p_T\to\infty$ \\
\cline{3-5}
Channel & $J$ & $gg$ & $gq$ & $q\bar{q}$ & $gg$ \\
\hline
${}^1S_0^{[8]}$ & $0$ & $1/p_T^2$ & $1/p_T$ & const     & $1/p_T^6$ \\
${}^3P_0^{[8]}$ & $0$ & $1/p_T^2$ & $1/p_T$ & $1/p_T^2$ & $1/p_T^6$ \\
${}^3P_2^{[8]}$ & $2$ & $1/p_T^2$ & $1/p_T$ & $1/p_T^2$ & $1/p_T^6$ \\
\hline
${}^3S_1^{[8]}$ & $1$ & const     & $1/p_T$ & $1/p_T^2$ & $1/p_T^4$ \\
${}^3P_1^{[8]}$ & $1$ & const     & $p_T$   & $1/p_T^2$ & $1/p_T^6$ \\
${}^3S_1^{[1]}$ & $1$ & const     & ---     & ---       & $1/p_T^8$ \\
\hline\hline
\end{tabular}
\caption{Power of $p_T$ in $d\hat{\sigma}/d\hat{t}$ for $p_T\to0$ and $p_T\to\infty$ at fixed $(y_3,y_4)$, from the squared amplitudes of Sec.~\ref{sec:amplitudes}; the horizontal rule separates the $J\neq1$ from the $J=1$ channels.  At large $p_T$ only the $gg$ subprocess is listed, since it dominates every channel there.  The origin of the entries is discussed in the text.}
\label{tab:powers}
\end{table}
 
This selection rule fixes the small-$p_T$ behavior of the $2\to2$ squared
amplitudes that enter our calculation.  Where a $2\to1$ process is open, the
additional parton of the $2\to2$ process may be soft or collinear and the squared
amplitude inherits the corresponding enhancement; where it is closed, that
enhancement is absent~\cite{Sun:2012vc}.
Expanding the amplitudes of Sec.~\ref{sec:amplitudes} at
fixed $(y_3,y_4)$, where $\hat{s}-M^2\to2p_TM\cosh\Delta_y$ and
$\hat{t},\hat{u}\to-p_TMe^{\mp\Delta_y}$, gives
$d\hat{\sigma}/d\hat{t}\propto p_T^{\,n}$ with the powers collected in
Table~\ref{tab:powers}.  The $gg$ column reproduces the grouping of
Fig.~\ref{fig:channel_breakdown} exactly: $n=-2$ for the three $J\neq1$ octets,
$n=0$ for the two $J=1$ octets and for the color singlet.  
The other columns do not sort by $J$: the $gq$ and $q\bar{q}$ powers, and the large-$p_T$ powers, each differ between channels of the same $J$ and coincide for channels of different $J$.  The $J$ organization of the $\cos\chi$ shape is therefore specifically a property of the $gg$ subprocess at small $p_T$; it carries over to the channel totals because $gg$ is the largest subprocess in every channel over the fiducial range.
Since $d^3\sigma/dy_3\,dy_4\,dp_T \propto p_T\,d\hat{\sigma}/d\hat{t}$, the $gg$
contribution to the $J\neq1$ channels grows as $1/p_T$ toward the collinear boundary
while that to the $J=1$ channels and to the singlet vanishes linearly there.

Two entries of Table~\ref{tab:powers} have other origins: the $q\bar q$ enhancement of the ${}^3P_J^{[8]}$ states comes from
soft-gluon emission off the $c\bar c$ pair, the term behind the ${}^3P_J^{[8]}\to{}^3S_1^{[8]}$ operator mixing in
NRQCD~\cite{Petrelli:1997ge}, and the $gq$ column stops at $1/p_T$ because the additional parton is a quark.

The last column of Table~\ref{tab:powers} is governed by a second, related rule:
at $p_T\to\infty$ what matters is whether the state can be formed by a single
off-shell gluon, whose $1^{--}$ quantum numbers select ${}^3S_1^{[8]}$ alone and
give it the leading-power $1/p_T^4$ fragmentation behavior~\cite{Braaten:1994vv}.  The
${}^3S_1^{[8]}$ channel therefore sits between the two rules: closed to two
on-shell gluons and so suppressed at small $p_T$ relative to the $J\neq1$ octets, open to one off-shell gluon and so the leading channel at large $p_T$.

Because $\cos\chi$ maps monotonically onto $p_T$ over most of the fiducial region,
the distribution interpolates between these two regimes---the backward peak
sampling the large-$p_T$ behavior, the forward region the small-$p_T$ one.  That
forward region is reached only at the smallest $p_T$ the acceptance admits: within
$|y_3|,|\eta_4|\leq1.1$ and $p_T\geq0.5\,\mathrm{GeV}$, $\cos\chi>0$ requires
$p_T\lesssim0.7\,\mathrm{GeV}$, as can be seen in Figs.~\ref{fig:coschi_map} and~\ref{fig:coschi_max}.
The grouping of Fig.~\ref{fig:channel_breakdown} follows: the channels with an
open two-gluon process sit higher and fall away more slowly as $\cos\chi$ increases toward the forward cutoff. (The sharp drop at the boundary itself is channel independent, being set by the acceptance alone; Appendix~\ref{app:coschi_max}.)

The subprocess decomposition of Fig.~\ref{fig:channel_breakdown} follows the same rule.  The $gg$ subprocess dominates the
$J\neq1$ channels---$gq$ and $q\bar q$ supply about $15\%$ across the forward half of the range at LHCb and under $10\%$ at sPHENIX---while in the $J=1$ octets,
where $gg$ loses the small-$p_T$ enhancement, the quark-initiated subprocesses carry a larger share: about a quarter of
${}^3S_1^{[8]}$ forward of $\cos\chi\approx-0.2$ at LHCb and about half at sPHENIX.  ${}^3S_1^{[8]}$ is the one channel whose open $2\to1$ process, $q\bar q\to{}^3S_1^{[8]}$, the $gq$
subprocess reaches through a $g\to q\bar q$ splitting on the incoming gluon.
 
The ${}^3P_J^{[8]}$ substates make the case for $J$
sharpest.  They share identical orbital and spin quantum numbers ($L=1$, $S=1$)
and their LDMEs are related by heavy-quark spin symmetry
(Sec.~\ref{sec:nrqcd}), so were the squared amplitude $J$-independent at the
level of the angular distribution, the three $d\Sigma^{{}^3P_J^{[8]}}/d\cos\chi$
would coincide up to the overall $(2J+1)$ multiplicity.
Figure~\ref{fig:channel_breakdown} shows that they do not, and
the reason is given in Table~\ref{tab:powers}: ${}^3P_1^{[8]}$ is the one substate whose
two-gluon process $gg\to c\bar{c}[n]$ is closed, so it groups with the $J=1$ $S$-wave
${}^3S_1^{[8]}$, while ${}^3P_{0,2}^{[8]}$ retain the plateau and group with the
spin-singlet ${}^1S_0^{[8]}$.  What organizes the shape is therefore neither $L$
nor $S$, common to all three substates, but the total angular momentum $J$,
acting through the two-gluon initial state.
As noted in Sec.~\ref{sec:nrqcd}, the three substates enter an LDME fit only through their fixed combination; the ${}^3P_J^{[8]}$ contrast demonstrates the mechanism rather than supplying an additional degree of freedom.
 
The color-singlet ${}^3S_1^{[1]}$ steepens toward the forward cutoff for the same
reason as the $J=1$ octets, and this forward behavior---the part of the shape that carries the $J$ signature---is explained by the selection rule.  
In the central acceptance it parts from the octet ${}^3S_1^{[8]}$ only in the mid-backward region, where it develops a
weak shoulder and ${}^3S_1^{[8]}$ does not; there $p_T\sim1$--$2\,\mathrm{GeV}$
and the asymptotic classification of Table~\ref{tab:powers} no longer applies, so
this remains a separate feature 
of the full matrix element that we have not reduced to a simple rule; it does not affect the $J$ classification, which rests on the forward behavior.

The QpTC shares the same channel grouping (Fig.~\ref{fig:qec_vs_qptc}): its weight differs from the QEC's only by the kinematic
factor of Eq.~\eqref{eq:qptc_qec_ratio_LO}, which modulates the magnitude but not the grouping in $\cos\chi$.
The enhanced discrimination at the wider acceptances motivates the LDME-scenario analysis of Sec.~\ref{sec:ldme}, where the central and forward acceptances are compared for their power to separate the three LDME sets.

\subsection{Dependence on the collinear cutoff $p_T^{\min}$}
\label{subsec:pt_tuning}
 
\begin{figure}
\centering
\includegraphics[width=\linewidth]{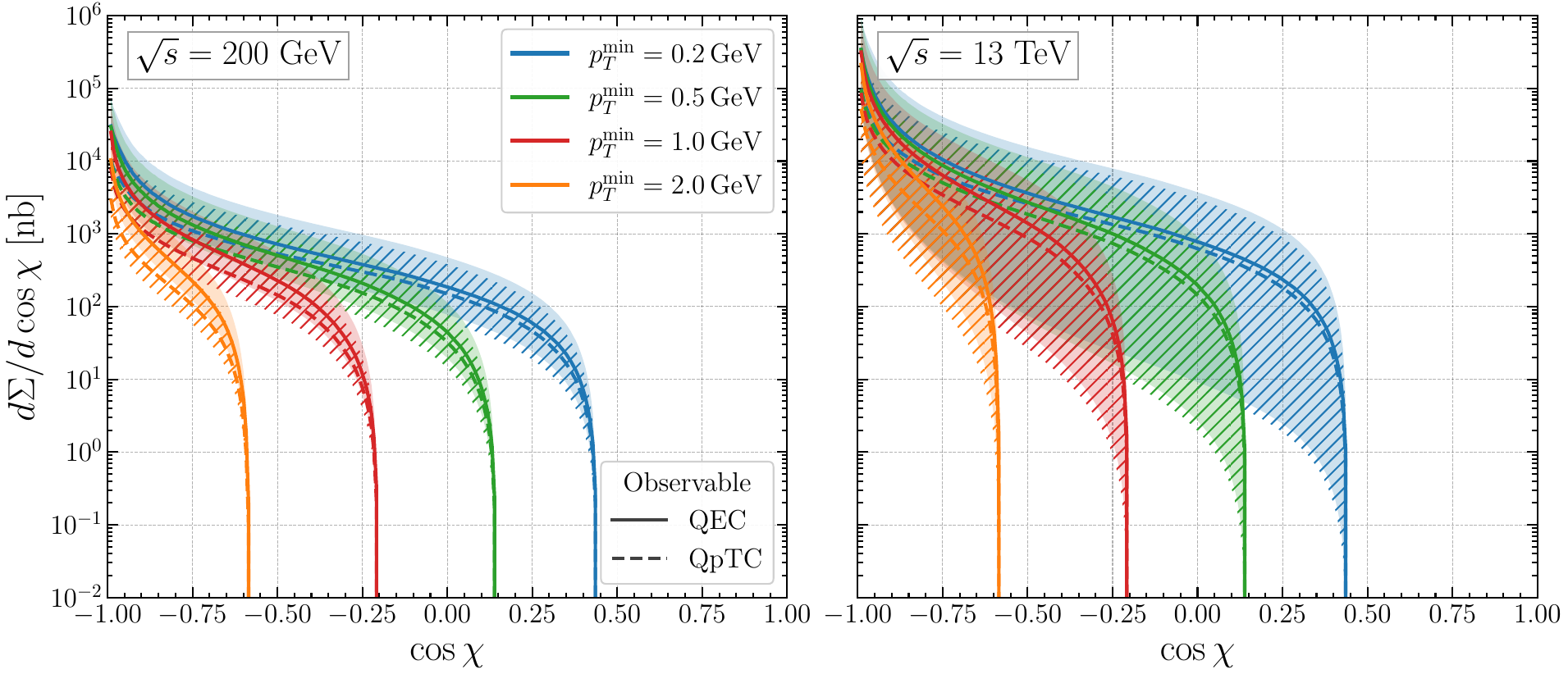}
\caption{LDME-weighted total QEC (solid, filled 7-point scale bands) and QpTC (dashed, hatched bands) with the Default LDMEs of Table~\ref{tab:ldme}, at $\sqrt{s}=200\,\mathrm{GeV}$ (left) and $13\,\mathrm{TeV}$ (right), both in the central acceptance $|y_3|,|y_4|\leq1.1$ of Eq.~\eqref{eq:event_cuts}.  At $13\,\mathrm{TeV}$ this rapidity range is narrower than the ATLAS/CMS acceptance of Fig.~\ref{fig:channels_atlascms} ($|y_3|\leq2$, $|\eta_4|\leq2.5$), hence the smaller $\cos\chi$ range.  The four values of the collinear cutoff $p_T^{\min}$ give $\cos\chi_{\max}^{\rm fid}\approx+0.44$, $+0.14$, $-0.20$, and $-0.58$ at both energies (Appendix~\ref{app:coschi_max}).}
\label{fig:pt_min_scan}
\end{figure}
 
So far the collinear cutoff has been fixed at $p_T^{\min}=0.5\,\mathrm{GeV}$.  Here we vary it: raising $p_T^{\min}$ removes the forward part of the leading-order distribution, leaving that region to soft and higher-order sources, shifts the weight toward the $J=1$ channels, and moves the calculation away from the low-$p_T$ region where fixed order is least reliable.
Figure~\ref{fig:pt_min_scan} illustrates the LDME-weighted total QEC and QpTC at $\sqrt{s}=200\,\mathrm{GeV}$ and $13\,\mathrm{TeV}$ for four values of $p_T^{\min}$.
 
At small $p_T^{\min}=0.2\,\mathrm{GeV}$, the kinematic boundary
$\cos\chi_{\max}^{\rm fid} \approx +0.44$ exposes the largest forward window
accessible within the fiducial acceptance.  The hard contribution
covers $\cos\chi \in [-1, +0.44]$ with channel-dependent shape
(Sec.~\ref{subsec:jstructure}). 
Across the backward region the non-perturbative soft contributions are subdominant in the QEC (Fig.~\ref{fig:soft_radiation}), and in Fig.~\ref{fig:pt_min_scan} the QpTC
(dashed) tracks the QEC with the kinematic suppression of
Eq.~\eqref{eq:qptc_qec_ratio_LO}.
 
At large $p_T^{\min}=2\,\mathrm{GeV}$, the forward boundary retracts
to $\cos\chi_{\max}^{\rm fid} \approx -0.58$ (Appendix~\ref{app:coschi_max}),
removing the entire forward hemisphere and a substantial portion of
the backward hemisphere from the kinematic acceptance.  The QEC is
then concentrated in $\cos\chi \in [-1,\,-0.58]$, sampling only the
hard backward region where the perturbative LO calculation is more
reliable (though, as Sec.~\ref{subsec:ldme_sensitivity} shows, not the region in which the LDME sets are best separated).
The excluded forward region receives no perturbative LO
contribution at all, so any signal there originates 
from sources beyond the leading-order hard process---the non-perturbative soft emission of Sec.~\ref{sec:soft}, higher-order radiation, feed-down decays, and the underlying event---a window for hadronization studies once the others are controlled, and
the same forward regime probed at high $J/\psi$ $p_T$ in
Ref.~\cite{Chen:2024nfl}.
 
The cutoff $p_T^{\min}$ is thus a kinematic knob on the soft sensitivity, independent of the weight: where the QpTC suppresses
the hard contribution by reweighting (Sec.~\ref{sec:soft}), raising $p_T^{\min}$ removes it from the forward phase space.  It
also shifts the channel balance: removing the small-$p_T$ region, in which only the $J\neq1$ channels carry the two-gluon
enhancement, raises the share of the $J=1$ channels (${}^3S_1^{[1]}$, ${}^3S_1^{[8]}$, ${}^3P_1^{[8]}$) from 8--12\% to 17--19\% of $\Sigma_{\rm QEC}$ between $p_T^{\min}=0.2$ and $2\,\mathrm{GeV}$ (central acceptance, $200\,\mathrm{GeV}$ and $13\,\mathrm{TeV}$).

Raising the cutoff also moves the calculation away from the region where fixed order is least reliable.  For $p_T\lesssim M$ the
transverse momentum is generated by soft and collinear radiation and a transverse-momentum-dependent treatment is
required~\cite{Berger:2004cc,Sun:2012vc,Echevarria:2019ynx,Fleming:2019pzj}; color-octet production there requires
gauge-completed matrix elements~\cite{Nayak:2005rt}, and TMD factorization is violated for $P$-wave states~\cite{Ma:2014oha}; and
at the LHC acceptances the momentum fractions reach $x\sim10^{-5}$, where small-$x$ effects have been invoked for low-$p_T$
$J/\psi$ production~\cite{Ma:2014mri}.  The $J\neq1$ enhancement of Sec.~\ref{subsec:jstructure} is the fixed-order form of that
radiation: which channels carry it is set by the open $2\to1$ processes, but its mapping onto $\cos\chi$ at the lowest $p_T$ is
modified by resummation and the parton shower.  The wider acceptances keep the forward region at higher $p_T$: at LHCb, over $-0.7\leq\cos\chi\leq0.14$, the $J=1$ octets fall by a factor of about 2.5 more than the $J\neq1$ octets, for $p_T^{\min}=0.5$ and $2\,\mathrm{GeV}$ alike (not shown).

% ===========================================================================
\section{LDME Sensitivity and Experimental Prospects}
\label{sec:ldme}
 
A central feature of NRQCD phenomenology at hadron colliders is that
the inclusive $J/\psi$ $p_T$ spectrum constrains only certain
linear combinations of the color-octet LDMEs, leaving the
individual values weakly
constrained~\cite{Cho:1995vh,Cho:1995ce,Ma:2010yw,Chao:2012iv}. The three LDME scenarios of Table~\ref{tab:ldme}---Default,
${}^1S_0^{[8]}$-min, and ${}^1S_0^{[8]}$-max---illustrate this
degeneracy: extracted from simultaneous fits to Tevatron CDF data 
(Sec.~\ref{sec:amplitudes}), all three are consistent with the ATLAS and CMS 
prompt $J/\psi$ cross section, up to $p_T = 70\,\mathrm{GeV}$~\cite{Chao:2012iv,ATLAS:2011aqv,CMS:2011rxs}, 
yet they differ qualitatively in their decomposition over the
color-octet sector.  The Default set is the best fit of
Ref.~\cite{Chao:2012iv}, while ${}^1S_0^{[8]}$-min and ${}^1S_0^{[8]}$-max are the two
extreme choices along the direction in LDME space that the fit leaves undetermined; even these
two extremes were found to remain close to the Tevatron cross section and polarization data~\cite{Chao:2012iv}.  
The Tevatron polarization data therefore do not
lift the degeneracy (Ref.~\cite{Chao:2012iv} noted that polarization at $p_T\gtrsim20\,\mathrm{GeV}$ could), and a measurement that resolves the difference
among these three sets would directly constrain the LDMEs individually
rather than the degenerate combinations.

\subsection{LDME sensitivity at leading order}
\label{subsec:ldme_sensitivity}
 
In general-purpose event generators the LDMEs enter through the leading-order hard processes---in
\textsc{Pythia\,8} the $2\to2$ matrix elements of Sec.~\ref{sec:amplitudes}, supplemented in recent
versions by NRQCD splittings in the shower~\cite{Cooke:2023ldt}, as also in \textsc{Herwig\,7}~\cite{Masouminia:2025kec}---so the LO correlator is where an LDME dependence of the QEC must first appear, and where its size relative to the perturbative
uncertainty can be judged.  
We therefore assess the LDME-discriminating power of the QEC at the analytic LO level, where each channel enters with the shape of its hard scattering alone, before the soft emission (Sec.~\ref{sec:soft}), parton shower, and hadronization of a full simulation are added.
The per-channel results of Sec.~\ref{sec:amplitudes}, computed with the
Default LDMEs, are re-weighted by the LDME ratios of
Table~\ref{tab:ldme}:
\begin{equation}
\label{eq:ldme_reweight}
\Sigma^{\rm scenario}_{\rm ch}
\;=\;
\Sigma^{\rm Default}_{\rm ch}
\;\times\;
\frac{\LDME{n}_{\rm scenario}}
     {\LDME{n}_{\rm Default}},
\end{equation}
where $\Sigma_{\rm ch}$ denotes the integrated or differential
observable for channel $n={}^{2S+1}L_J^{[c]}$. The LDME-weighted total is then obtained by summing over the six channels of Eq.~\eqref{eq:total_qec_qptc}, whose weights are set by the three independent octet LDMEs of Table~\ref{tab:ldme}, the singlet being fixed.
 
\begin{figure}
\centering
\includegraphics[width=\linewidth]{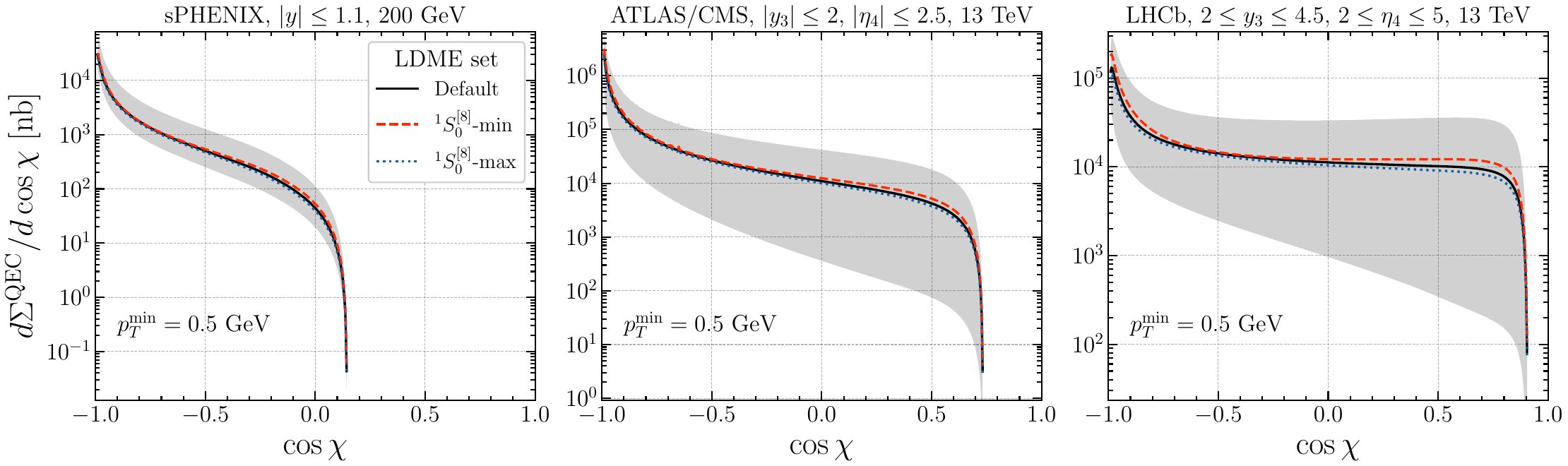}
\caption{Absolute LDME-weighted QEC for the three sets of Table~\ref{tab:ldme} at sPHENIX ($\sqrt{s}=200\,\mathrm{GeV}$, left), ATLAS/CMS ($13\,\mathrm{TeV}$, center), and LHCb ($13\,\mathrm{TeV}$, right), each with its 7-point scale band.  At every acceptance the three sets overlap within the leading-order bands; the normalized distributions are compared in Fig.~\ref{fig:ldme_ratio_qec}.}
\label{fig:ldme_comparison_qec}
\end{figure}

The overall rate cannot separate the scenarios.  Summed over all channels, the integrated QEC differs by
16--39\% across the three
sets (the sets exchange ${}^1S_0^{[8]}$ for ${}^3S_1^{[8]}$ and ${}^3P_J^{[8]}$ along the direction the $p_T$ fits leave free, which largely preserves the rate), well below the leading-order scale
uncertainty on the rate itself, a factor of two or more in each direction: at this order the
QEC is proportional to $\alpha_s^3$ and to the parton densities, with no
higher-order compensation. The three absolute
distributions therefore overlap within their scale bands
(Fig.~\ref{fig:ldme_comparison_qec}) and cannot tell the LDME sets apart.

\begin{figure}
\centering
\includegraphics[width=\linewidth]{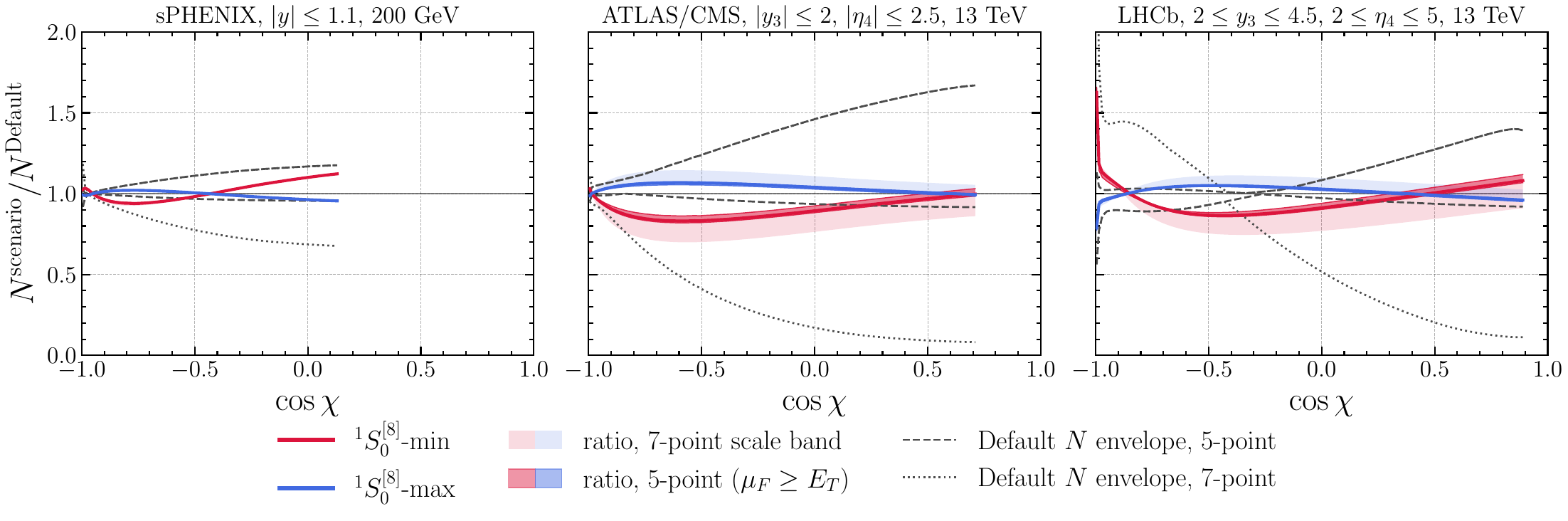}
\caption{Normalized QEC, $N=\Sigma^{-1}d\Sigma/d\cos\chi$ ($\Sigma$ over the full fiducial range), of the ${}^1S_0^{[8]}$-min (red) and ${}^1S_0^{[8]}$-max (blue) sets divided by the Default, at the three acceptances of Fig.~\ref{fig:ldme_comparison_qec}. 
Bands: the ratio formed at a common scale, over the 7-point variation (light) and without $\mu_F=E_T/2$ (dark); at sPHENIX both are narrower than the lines.
Gray lines: envelope of the Default's own $N$ relative to its central value, over the same seven (dotted) and five (dashed) scale choices.  Over most of the range the two share their upper edge (at LHCb below $\cos\chi\approx-0.5$, their lower edge); the difference comes from $\mu_F=E_T/2\approx1.5$~GeV, where the small-$x$ gluon grows steeply with the scale.}
\label{fig:ldme_ratio_qec}
\end{figure}

Since the rate is not predicted reliably at this order, the quantity to confront with data is the normalized
distribution $N=\Sigma^{-1}\,d\Sigma/d\cos\chi$, with $\Sigma$ integrated over the full fiducial range.  Normalizing
removes the common factors---the $\alpha_s^3$ of the matrix elements and the overall parton luminosity---that make the
rate uncertain by a factor of two or more, and leaves the scale dependence of the shape. 
The normalized shape depends on the LDME set (Fig.~\ref{fig:ldme_ratio_qec}): relative to the Default, the ${}^1S_0^{[8]}$-min set produces a dip of 13--17\% around $\cos\chi\approx-0.5$ relative to both ends (6\% near $-0.8$ at sPHENIX), and the ${}^1S_0^{[8]}$-max set the opposite with 35--40\% of that amplitude.
The bend reflects the channels the sets exchange: relative to the total, the normalized ${}^1S_0^{[8]}$ distribution is nearly
flat, ${}^3S_1^{[8]}$ is concentrated toward $\cos\chi=-1$, and $\sum_J{}^3P_J^{[8]}$ rises toward the forward end, so trading
${}^1S_0^{[8]}$ for the other two octets raises both ends against the middle.  
In the ratio to the Default formed at a common scale, most of the scale dependence cancels: 
the pattern is stable over the scale variation without $\mu_F=E_T/2$ (dark bands in Fig.~\ref{fig:ldme_ratio_qec}), and that lowest scale enlarges it at the LHC acceptances (light bands). This cancellation does not extend to the normalized distribution itself, which is what a measurement is compared with.

We note that the degeneracy of the three sets of Table~\ref{tab:ldme} relies on the next-to-leading-order short-distance coefficients, in which the
${}^3P_J^{[8]}$ contribution becomes negative at large $p_T$ and nearly
cancels that of ${}^3S_1^{[8]}$. All three sets satisfy
$\langle \mathcal{O}^{J/\psi}({}^3S_1^{[8]})\rangle
- 0.56\,\langle \mathcal{O}^{J/\psi}({}^3P_0^{[8]})\rangle/m_c^2 \simeq 0$~\cite{Chao:2012iv}.
At leading order the ${}^3P_J^{[8]}$ short-distance coefficient is
positive and this cancellation does not occur. The size of the bend at leading order should therefore be interpreted
with care, since the LDMEs were extracted with next-to-leading-order
short-distance coefficients. The effect is expected to be largest near
$\cos\chi =-1$ in the LHCb acceptance, where ${}^3S_1^{[8]}$ is concentrated.

The shape of the normalized distribution itself, however, still changes with the scale at leading order: over the 7-point variation the half-width of its band, relative to the central value, is 22\%,
68\% and 39\% at the three acceptances (sPHENIX, ATLAS/CMS, LHCb; median for $\cos\chi>-0.5$; dotted gray lines in Fig.~\ref{fig:ldme_ratio_qec}).
This comes mostly from  $\mu_F$ at its lowest value, $E_T/2\approx1.5$~GeV, where the small-$x$ gluon evolves rapidly and nothing compensates the dependence at this order.  
Without that point the half-width is 10\%, 29\% and 10\% (dashed gray lines in Fig.~\ref{fig:ldme_ratio_qec})---several times smaller than the uncertainty of the rate, but comparable to the LDME pattern.  
This is a property of leading-order production at low $p_T$ rather than of the QEC: for the normalized $J/\psi$
$p_T$ spectrum of the same events and the same scale variation, the LDME pattern is smaller relative to the band by a factor of up
to two at the LHC acceptances (1.8--2.0 for the 5-point band, 1.1--2.0 for the 7-point one), and larger by 1.6--1.8 at
sPHENIX.\footnote{We divide the cumulative Default distribution at the central scale ($\Sigma$ for the QEC, $\sigma$ for the
$p_T$ spectrum) into 20 equal parts and, in these bins, compare the rms of $\ln(N_S/N_{\rm Default})$ with the rms
half-width of the scale band of $\ln N_{\rm Default}$; the comparison changes little for 10--80 bins.}
The QpTC behaves like the QEC.

At leading order the QEC thus identifies the LDME signature---a bend of the normalized distribution around $\cos\chi\approx-0.5$---but
does not resolve it: the signature is smaller than the uncompensated $\mu_F$ dependence of the leading-order shape, and comparable to it without $\mu_F=E_T/2$.  A calculation beyond leading order (Sec.~\ref{sec:conclusion}), together with the full-final-state work, is what a quantitative extraction requires (Secs.~\ref{subsec:ldme_implications} and~\ref{sec:conclusion}).

\subsection{Experimental prospects}
\label{subsec:ldme_implications}

Within leading order, our analysis gives a clear assessment of the discriminating power of the QEC and QpTC: the overall rate is not usable,
and the LDME signature in the normalized shape is at best comparable to its leading-order scale uncertainty (Sec.~\ref{subsec:ldme_sensitivity}).
At sPHENIX the forward end of that bend is limited in three ways: it is a small part of the cross section, it sits directly above the collinear
cutoff, and the soft emission of Sec.~\ref{sec:soft} overtakes the hard contribution already at $\cos\chi\approx-0.2$
(Fig.~\ref{fig:soft_radiation}), so that the $+12\%/-5\%$ change of the normalized distribution at the forward edge ($\cos\chi\approx0.14$ in Fig.~\ref{fig:ldme_ratio_qec}), a property of the hard component alone, is diluted in the measured distribution by an isotropic soft contribution that carries only the summed octet rate.  
The first two limitations are lifted by the wider forward acceptances of Sec.~\ref{subsec:wide_rapidity}, which enlarge the region and move it clear of the cutoff.  Repeating the two-body model of Sec.~\ref{sec:soft} at the LHC acceptances shows that, for the QEC, the third limitation is lifted as well.  At LHCb $30\%$ of the hard correlator lies forward of $\cos\chi=0$, while the soft contribution, $8\%$ of the hard QEC in total, adds only $17\%$ to it in this region and overtakes the hard contribution only above $\cos\chi\approx0.9$, so the forward window of the QEC is set by the hard channel shapes. For the QpTC the soft term is relatively larger ($34\%$ in total, $80\%$ of the hard term forward of $\cos\chi=0$) and dilutes the hard channel shapes more strongly. At ATLAS and CMS the forward hemisphere holds only $2\%$ (QEC) and $5\%$ (QpTC) of the hard contribution, and the soft contribution overtakes the hard one above $\cos\chi\approx0.6$ and $0$, respectively.

One caveat applies specifically to the forward window.  The \textsc{Pythia\,8} study
of Ref.~\cite{Zhang:2026tjo} finds that the transition from the parton to the hadron
level suppresses the forward correlator by about an order of magnitude, for $p_T>7\,\mathrm{GeV}$,
charged hadrons and the $p_T$ weighting, so the size of the effect does not transfer directly to our
kinematics.  A suppression common to all channels cancels in the scenario ratios of
Fig.~\ref{fig:ldme_ratio_qec}, but changes the normalized distribution compared with data unless it is
independent of $\cos\chi$; how it depends on $\cos\chi$ and on the channel only the full-final-state
simulation left to future work can settle. 
In the kinematic window of the preliminary STAR measurement mentioned in Sec.~\ref{sec:intro},
the leading-order recoil is confined to $\cos\chi \lesssim -0.9$, so the
correlator at $\cos\chi>0$ lies entirely beyond the reach of the
$2\to2$ hard process and is determined by the sources discussed above.
 
A useful experimental program can be built on these observables.
The discriminating handle is the shape of the normalized distribution, notably its bend around $\cos\chi\approx-0.5$; the energy-flow measurement, moreover, carries
systematics distinct from those of the inclusive $p_T$ spectrum.
A combined fit of the QEC (and QpTC) shape with the inclusive
$p_T$ spectrum can leverage these partly independent systematics
to tighten constraints on the LDME combinations probed by the
data, even when no single observable by itself breaks the
degeneracy.  Combination with LHC measurements and
$J/\psi$ polarization data---which constrain the spin-triplet 
octet LDMEs $\LDME{{}^3S_1^{[8]}}$ and $\LDME{{}^3P_J^{[8]}}$ 
through angular distributions independent of the unpolarized 
$\cos\chi$ studied here---is essential to make further progress.

Experimentally, the sPHENIX tracking and calorimetry cover $|\eta|\leq1.1$~\cite{Belmont:2023fau}, which motivates the rapidity acceptance used here.
The $p_T$- and rapidity-differential $J/\psi$ cross sections have already been measured at RHIC at the same $\sqrt{s}$ and at central rapidity~\cite{PHENIX:2011gyb,STAR:2012wnc,STAR:2018smh}, and leading-order calculations can be compared with them.
A major experimental challenge for
both observables is the subtraction of the underlying event, which adds substantial activity on top of the hard scatter.
Resolving the $\cos\chi$ distribution to the precision required for
the LDME discrimination identified here will require careful control
of underlying-event and detector-unfolding systematics.

A further experimental caveat concerns feed-down.  A measured prompt
$J/\psi$ sample also contains $J/\psi$ from $\psi(2S)\to J/\psi+X$ and
$\chi_{cJ}\to J/\psi+\gamma$, which cannot be separated event by event.
Our normalization is consistent with such a sample: the LDMEs of
Table~\ref{tab:ldme} were fitted to prompt $J/\psi$ data with direct-production short-distance coefficients~\cite{Chao:2012iv}, and so absorb the feed-down contributions effectively.
The QEC itself, however, is sensitive to the decay products.  The accompanying
pions or photon carry ${\sim}0.2$--$0.4\,\mathrm{GeV}$ in the parent rest
frame---comparable to the soft-gluon energy of
Eq.~\eqref{eq:soft_decay}---and enter the energy sum of
Eq.~\eqref{eq:qec_def} at an angle uncorrelated with the hard recoil,
populating the same forward region as the color-octet soft emission of
Sec.~\ref{sec:soft}.  Quantifying this contamination, and whether the
measured $\psi(2S)$ and $\chi_{cJ}$ yields suffice to constrain it,
requires a full-final-state simulation, which we leave to future work.

% ===========================================================================
\section{Conclusions}
\label{sec:conclusion}

We have presented the first complete leading-order analytic calculation
of the quarkonium energy correlator in $pp$ collisions, covering all
$2\to2$ partonic processes and treating the color-singlet
${}^3S_1^{[1]}$ and the leading color-octet NRQCD states on an equal
footing.  Where the hadron-collider estimate of Ref.~\cite{Chen:2024nfl}
was built from the quarkonium fragmentation functions alone, omitting the
recoil parton, we compute that contribution from the full hard matrix
elements, without expanding in $M/p_T$.  The analytic formula,
Eq.~\eqref{eq:qec_analytic}, resolves the non-monotonic, multi-valued map
between the recoil transverse momentum and $\cos\chi$; it reproduces the
\textsc{HELAC-onia} tree-level Monte Carlo in both shape and absolute
normalization and carries over, with the per-event weight replaced, to the lab-frame $p_T$-weighted variant (QpTC).

The analytic solution is the hard, leading-order baseline on which parton showering, the underlying event and hadronization act.
In particular, it fixes the boundary $\cos\chi_{\max}^{\rm fid}$ beyond which the $2\to2$ recoil cannot contribute, as a function of
$p_T^{\min}$ and the rapidity acceptance.  Beyond this boundary the remaining sources are mainly: higher-order real emission, the
non-perturbative soft emission from color-octet hadronization (more prominent in the QpTC), and the feed-down and underlying-event
contributions of Sec.~\ref{subsec:ldme_implications}.  At high $p_T$ this is the forward regime studied in Ref.~\cite{Chen:2024nfl}.
Since the boundary moves with the collinear cutoff $p_T^{\min}$ and with the acceptance, these two choices set the balance between hard and soft
sensitivity in a measurement.

Even where the hard interaction is the signal rather than a background,
the leading-order shape is informative.  The forward fall-off of the QEC
is organized by the total angular momentum $J$ of the intermediate
$c\bar{c}$ pair: the $J=1$ octets fall steeply, the $J\neq1$ octets stay flat much further forward, and the color-singlet ${}^3S_1^{[1]}$ crosses
between the two as the acceptance widens. 
The mechanism is the
selection rule that closes $gg\to c\bar{c}[n]$ for $J=1$, and with it the
small-$p_T$ enhancement the $J\neq1$ channels carry~\cite{Sun:2012vc}; the forward region,
reached at low $p_T$, displays that
difference. 
In this sense the $\cos\chi$ distribution is the
rest-frame image of the familiar $p_T$ hierarchy of the NRQCD channels, with the
acceptance and the collinear cutoff selecting which part of that hierarchy is on
display (Table~\ref{tab:powers}).  
The LDMEs are fitted to the inclusive spectrum at $p_T\gtrsim7\,\mathrm{GeV}$, where the channels are ordered by their large-$p_T$
behavior, set by fragmentation.  
The ordering by $J$---rather than by $L$ or $S$ separately, as the three ${}^3P_J^{[8]}$ substates
show---is instead a small-$p_T$ effect, and the QEC maps small $p_T$ into its forward region.
At this order the correlator is a weighted projection of the $2\to2$ cross section.  Two properties nonetheless make it useful: in the $J/\psi$ rest frame the hard contribution is confined below $\cos\chi_{\max}^{\rm fid}$ while the soft emission is not, and the
observable is defined for any final state, without reference to a jet or a recoil.

Widening the rapidity acceptance opens the forward region in which the channels separate, from the compressed sPHENIX window to the well-separated forward reach of LHCb.  
For the LDMEs, however, the QEC inherits the degeneracy of the inclusive $p_T$ spectrum: the three sets of Table~\ref{tab:ldme}, which the $p_T$ fits do not distinguish, give integrated  correlators that differ by 16--39\%, well within the leading-order scale uncertainty of the rate. 
We have therefore examined the normalized distribution at the sPHENIX, ATLAS/CMS and LHCb acceptances.  The different LDME scenarios we considered do leave a signature in it, a bend of 13--17\% around $\cos\chi\approx-0.5$ at the LHC acceptances (6\% near $\cos\chi\approx-0.8$ at sPHENIX), but at leading order it is smaller than the factorization-scale uncertainty of the shape and comparable to it without the lowest scale---a limitation it shares with the $p_T$ spectrum of the same events, which fares worse at the LHC acceptances and better at sPHENIX.
At this order the QEC alone does not separate the LDME sets.  Its shape carries experimental systematics distinct from those of the $p_T$ spectrum, and it is in a simultaneous fit with the $p_T$ spectrum and the $J/\psi$ polarization, once the scale uncertainty of the shape has been reduced beyond leading order, that it can constrain the color-octet LDMEs.

Two developments would improve our ability to use the QECs as a quantitative discriminant of $J/\psi$ production channels.  For the LDMEs the decisive one is a calculation
beyond leading order.  A next-to-leading-order correlator, for which the machinery of NLO
$J/\psi$ $p_T$ calculations can be carried over, reduces the $\mu_F$ dependence that at
leading order is as large as the LDME pattern or larger and stabilizes the absolute normalization.  Since half of the correlator comes from $p_T\lesssim 1.4$--$2.2$~GeV, it has to be matched to a resummed or small-$x$
treatment of the low-$p_T$ region~\cite{Sun:2012vc,Echevarria:2019ynx,Fleming:2019pzj,Ma:2014mri},
or applied above a higher $p_T$ cut at the price of the forward window, which the wider
acceptances keep longest (Sec.~\ref{subsec:pt_tuning}).  The second is the full final state: a
study with a general-purpose event generator such as \textsc{Pythia\,8}, benchmarked against
the analytic baseline, must show whether the $J$-organized shape and the soft-radiation
sensitivity survive showering---in particular initial-state radiation at low $p_T$, which at fixed
order is the very enhancement that orders the channels---multi-parton interactions and hadronization, and quantify the feed-down from $\psi(2S)$ and $\chi_{cJ}$ decays that populates the forward region; with such a
sample, a machine-learning discriminant using the complete event would bound how much channel
separation is available at all.  The analytic baseline itself is not tied to the QEC: it gives, event by
event, the LO probabilities of the intermediate states, from which observables tailored to the LDMEs---up
to the optimal observable of the channel likelihood---can be built and tested against the same scale
variation.  Applications to cold-nuclear-matter effects in $p+\mathrm{Au}$ and to the quark-gluon plasma in $\mathrm{Au}+\mathrm{Au}$
at sPHENIX, and joint analyses with $J/\psi$ polarization, are natural
further steps.

\begin{acknowledgments}
This work was supported by the U.S.~Department of Energy, Office of Science,
through the Office of Nuclear Physics, and by the LDRD Program at Los Alamos
National Laboratory under projects 20230857PRD2, 20230872PRD4 and 20250214ER.
LANL is operated by Triad National Security, LLC, for the National Nuclear
Security Administration of the U.S.~Department of Energy under Contract
No.~89233218CNA000001.
\end{acknowledgments}

\section*{Data availability statement}
There are no publicly available research data or software supporting this manuscript. Requests for further information or data should be sent to the authors.

% ===========================================================================
\appendix

\section{Closed-form inversion of \texorpdfstring{$\cos\chi(p_T)$}{cos chi(pT)}}
\label{app:ptsol}

The roots $p_T^{\rm sol}(\cos\chi)$ needed in Eq.~\eqref{eq:qec_analytic}
are not the solutions of a transcendental equation: after rationalization,
Eq.~\eqref{eq:coschi} reduces to a quartic in $p_T^2$, whose roots are
given in closed form by Ferrari's method~\cite{Abramowitz:1964}.
Throughout this appendix, we abbreviate $s_\chi \equiv \sin\chi$ and $w \equiv p_T^2$.

\subsection{Reduction to a quartic}

Equation~\eqref{eq:coschi} is already written in a form containing only the
two independent radicals $E_T \equiv \sqrt{p_T^2 + M^2}$ and $|\vec p_3|$
[Eq.~\eqref{eq:p3mag}], both with radicands polynomial in $p_T^2$. Writing the
constraint $\cos\chi(p_T)=\cos\chi$ as
\begin{equation}
\cos\chi\,|\vec p_3|\,(P_2 + P_3\,E_T) = -\,(P_0 + P_1\,E_T),
\label{eq:grouped}
\end{equation}
with
\begin{equation}
\begin{gathered}
P_0 = p_T^2\cosh y_3\cosh\Delta_y + M^2\sinh y_3\sinh\Delta_y ,\\
P_1 = p_T\cosh y_3 ,\quad
P_2 = p_T ,\quad
P_3 = \cosh\Delta_y ,
\end{gathered}
\end{equation}
and squaring twice---first to remove $|\vec p_3|$, then $E_T$---eliminates
both radicals. The resulting polynomial in $p_T$ contains only even powers, so
the constraint is equivalent to the quartic $\sum_{k=0}^{4} a_k\,w^k=0$ of Eq.~\eqref{eq:quartic}, whose coefficients $a_k$ are elementary functions of $(y_3,y_4,\cos\chi)$ and $M$.  The leading and constant coefficients factor into transparent kinematic combinations,
\begin{align}
a_4 &= \cosh^4\!y_3\,\sinh^4\!\Delta_y\;s_\chi^4 ,
\label{eq:a4}\\
a_0 &= M^8\,\sinh^4\!y_3\,\bigl[(1-s_\chi^2) - s_\chi^2\sinh^2\!\Delta_y\bigr]^2 ,
\label{eq:a0}
\end{align}
while the intermediate coefficients, grouped by powers of $s_\chi^2$ and written
with the shorthand
$Q_1 \equiv \cosh^{2}\Delta_y\,\cosh^{2} y_3 + \sinh^{2}\Delta_y\,\sinh^{2} y_3$,
are
\begin{align}
a_3 &= 2 M^{2} \cosh^{2} y_3 \Big[\, s_\chi^{4}\, \sinh^{2}\Delta_y\; Q_1
     - s_\chi^{2} \big( (\cosh^{2}\Delta_y + 1)(\sinh^{2} y_3 + \sinh^{2} y_4)
     \nonumber \\
     &\qquad\qquad\qquad\quad
     - 4 \cosh\Delta_y\, \sinh y_3 \sinh y_4 \big) \Big],
\label{eq:a3}\\[4pt]
a_2 &= M^{4} \Big[\, s_\chi^{4}\big( \cosh^{4}\Delta_y
       + 2\cosh^{2}\Delta_y\,\sinh^{2} y_3\,(3\sinh^{2}\Delta_y + 1)
       \nonumber \\
       &\qquad\qquad\quad
       + \sinh^{4} y_3\,(6\sinh^{4}\Delta_y + 6\sinh^{2}\Delta_y + 1) \big) 
       \nonumber\\
    &\quad\quad\quad 
    - 2 s_\chi^{2}\big( \cosh^{2}\Delta_y\,\cosh^{2} y_3\,(2\sinh^{2} y_3 + \sinh^{2} y_4)
   \nonumber\\
    &\qquad\qquad\qquad 
       + \cosh^{2}\Delta_y\,\sinh^{2} y_3\,(\sinh^{2} y_3 + \sinh^{2} y_4) \nonumber\\
    &\qquad\qquad\qquad  
    - 4\cosh\Delta_y\,\sinh y_3 \sinh y_4\,(\cosh^{2} y_3 + \sinh^{2} y_3)
       + \cosh^{2} y_3\,\sinh^{2} y_3 
    \nonumber\\
    &\qquad\qquad\qquad 
       + \sinh^{4} y_3 + \sinh^{2} y_3\,\sinh^{2} y_4 \big) 
       \nonumber\\
    &\quad\quad\quad        
       + (\sinh^{2} y_3 - \sinh^{2} y_4)^{2} \Big],
\label{eq:a2}\\[4pt]
a_1 &= 2 M^{6} \sinh^{2} y_3 \Big[\, s_\chi^{4}\, \cosh^{2}\Delta_y\; Q_1
     - s_\chi^{2}\big( \cosh^{2}\Delta_y\,(\cosh^{2} y_3 + \sinh^{2} y_3 + \sinh^{2} y_4)
        \nonumber\\
    &\qquad\qquad
    + \sinh^{2} y_3\,(\sinh^{2}\Delta_y + 2)
    - 4\cosh\Delta_y\,\sinh y_3 \sinh y_4 \big)
     + (\sinh^{2} y_3 - \sinh^{2} y_4) \Big].
\label{eq:a1}
\end{align}

\subsection{Solution}

The quartic is solved in radicals by Ferrari's method---reduction to a depressed
quartic, a resolvent cubic solved by Cardano's formula, and factorization into two
quadratics~\cite{Abramowitz:1964}---which establishes that the inversion is closed-form.

Writing the quartic in monic form, $w^4 + b_3 w^3 + b_2 w^2 + b_1 w + b_0 = 0$
with $b_k = a_k/a_4$, and setting $w = t - b_3/4$ removes the cubic term:
\begin{equation}
t^4 + p\,t^2 + q\,t + r = 0,
\end{equation}
with
\begin{subequations}
\begin{align}
p &= b_2 - \tfrac{3}{8}b_3^2,\\
q &= b_1 - \tfrac{1}{2}b_3 b_2 + \tfrac{1}{8}b_3^3,\\
r &= b_0 - \tfrac{1}{4}b_3 b_1 + \tfrac{1}{16}b_3^2 b_2 - \tfrac{3}{256}b_3^4 .
\end{align}
\end{subequations}
A quartic with no cubic term factorizes into two quadratics with opposite linear coefficients, $(t^2-R\,t+\alpha)(t^2+R\,t+\beta)$; matching powers of $t$ gives $\alpha+\beta-R^2=p$, $R(\alpha-\beta)=q$, and $\alpha\beta=r$. With $m\equiv(\alpha+\beta)/2$ the first two fix $R^2=2m-p$ and $\alpha,\beta=m\pm q/(2R)$, and the third becomes the resolvent cubic
\begin{equation}
8 m^3 - 4 p\, m^2 - 8 r\, m + \bigl(4 p r - q^2\bigr) = 0 ,
\label{eq:resolvent}
\end{equation}
solved by Cardano's formula~\cite{Abramowitz:1964}; any root with $2m-p\neq0$ may be used. With $R \equiv \sqrt{2m-p}$ the two quadratic factors are
\begin{equation}
t^2 \mp R\,t + \Bigl(m \pm \frac{q}{2R}\Bigr) = 0 ,
\label{eq:twoquad}
\end{equation}
and their four roots give $w = t - b_3/4$ and $p_T^{\rm sol} = \sqrt{w}$, in radicals of the $a_k$.

Squaring Eq.~\eqref{eq:grouped} twice admits the sign combinations
$(\pm E_T,\pm|\vec p_3|)$, so the quartic $\sum_{k=0}^{4} a_k\,w^k=0$
[Eq.~\eqref{eq:quartic}] carries spurious
roots. The physical solutions are the $w$ that (i)~are real and positive, and (ii)~reproduce the input value on back-substitution into
Eq.~\eqref{eq:coschi}; condition~(ii) discards the roots introduced by
squaring. Outside the fold region of Sec.~\ref{subsec:helac} a single root survives, and the
sum in Eq.~\eqref{eq:qec_analytic} runs over the retained values.

\section{The maximum of $\cos\chi$ within fiducial acceptance}
\label{app:coschi_max}
 
In this appendix we derive the upper limit of $\cos\chi$ accessible
under the fiducial cuts of Sec.~\ref{subsec:analytic_QEC}.  The
$J/\psi$ rapidity $y_3$ and the recoil-parton pseudo-rapidity $\eta_4$
are given independent ranges, so we maximize Eq.~\eqref{eq:coschi} over
the box $y_3\in[y_3^{\min},y_3^{\max}]$,
$y_4\in[y_4^{\min},y_4^{\max}]$, $p_T\geq p_T^{\min}$ (at leading order
the recoil is massless, so $\eta_4=y_4$).  The three experimental
acceptances of this paper---sPHENIX and ATLAS/CMS, central and
symmetric in each rapidity, and LHCb, forward and asymmetric---are the
cases of interest.
 
\textit{Reduction to the $p_T=p_T^{\min}$ surface.}
The forward maximum of $\cos\chi$ is approached at small $p_T$ in the
same-sign branch of Eq.~\eqref{eq:coschi-pT0-limit}, in which $y_3$ and $y_4$ have the same sign and $|\Delta_y|$ is large.
Along that
branch $\cos\chi$ falls monotonically from its $p_T\to0$ value toward
$\cos\chi(p_T\to\infty)\to-1$, as we verify numerically for all three
acceptances, so its maximum over $p_T$ is attained at the lower boundary
$p_T=p_T^{\min}$.  The non-monotonic local maxima visible in Fig.~\ref{fig:coschi_pt} occur only for the
opposite-sign configurations, which remain in the backward
hemisphere and so do not affect the forward maximum.  Setting $p_T=p_T^{\min}$,
Eq.~\eqref{eq:coschi} becomes
\begin{align}
\label{eq:coschi-pTmin}
\cos\chi(y_3, y_4, p_T^{\min})
&= -
  \frac{p_T^{\min}\cosh y_3
       \bigl(E_T^{\min} + p_T^{\min}\cosh\Delta_y\bigr)
       + M^2 \sinh y_3 \sinh\Delta_y}
  {\bigl(p_T^{\min}+E_T^{\min}\cosh\Delta_y\bigr)\,
   \sqrt{(p_T^{\min})^2\cosh^2 y_3 + M^2\sinh^2 y_3}},
\end{align}
where $E_T^{\min}=\sqrt{M^2+(p_T^{\min})^2}$ and $\Delta_y=y_3-y_4$.
 
\textit{Location of the maximum.}
The same-sign requirement of the $p_T\to0$ limit
[Eq.~\eqref{eq:coschi-pT0-limit}] places the maximum on the edge
$y_4=y_4^{\max}$ (or its mirror for symmetric boxes), which maximizes $|\Delta_y|$. 
Where it sits along that edge depends on the acceptance
geometry.  For a symmetric acceptance (sPHENIX, ATLAS/CMS) the
stationarity condition $\partial\cos\chi/\partial y_3=0$ has an interior
solution, so the maximum is an interior stationary point in $y_3$ on the
$y_4$ edge.  For a forward-only acceptance such as LHCb's, that
stationary point falls outside the acceptance and the maximum migrates
to the corner $(y_3^{\min},y_4^{\max})$.  We locate it
numerically by maximizing $\cos\chi$ over the box---checking its interior,
edges, and corners---and cross-check against a brute-force grid scan.
 
\begin{figure}
\centering
\includegraphics[width=0.6\linewidth]{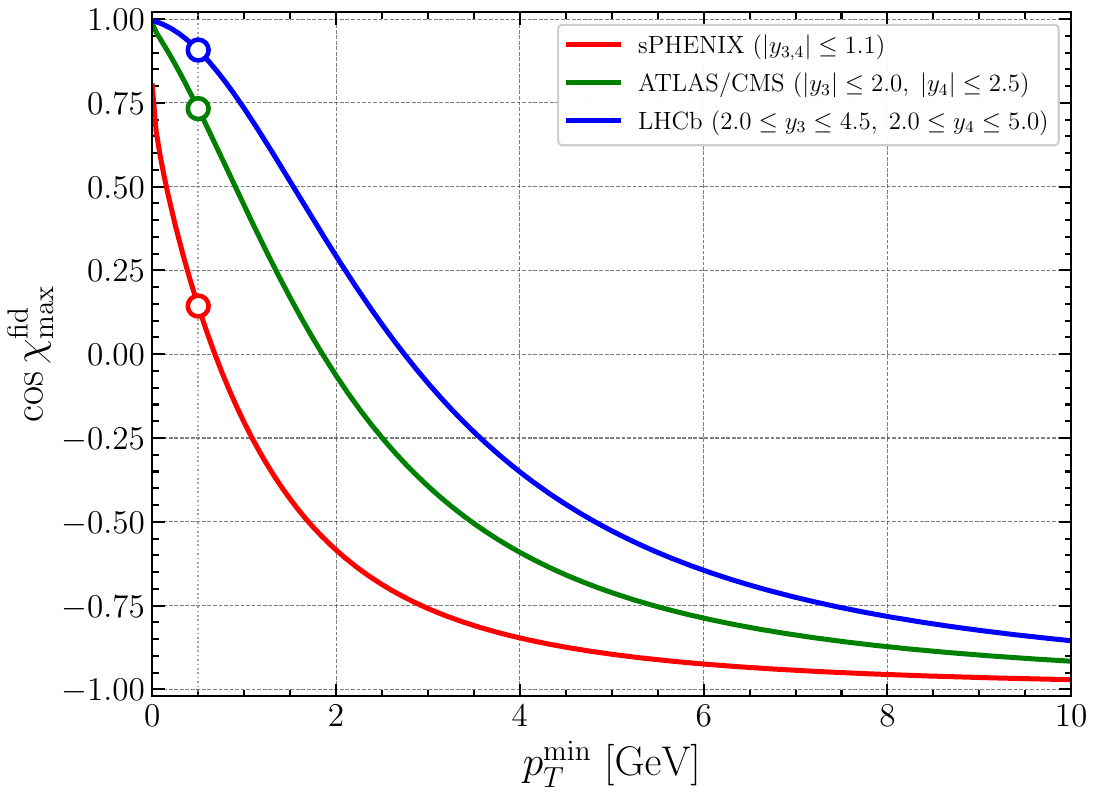}
\caption{Upper limit $\cos\chi_{\max}^{\rm fid}$ of the QEC within the three fiducial acceptances as a function of $p_T^{\min}$, from Eq.~\eqref{eq:coschi-pTmin}; open markers: the working point $p_T^{\min}=0.5\,\mathrm{GeV}$ ($\cos\chi_{\max}^{\rm fid}\approx0.14$, $0.73$, $0.91$).}
\label{fig:coschi_max}
\end{figure}

\textit{Numerical results.}
Figure~\ref{fig:coschi_max} shows $\cos\chi_{\max}^{\rm fid}$ as a
function of $p_T^{\min}$ for the three fiducial acceptances at
$M=3.0\,\mathrm{GeV}$.  At $p_T^{\min}=0.5\,\mathrm{GeV}$,
\begin{equation}
\label{eq:coschi-max-fid}
\cos\chi_{\max}^{\rm fid}\;\simeq\;
\begin{cases}
0.14 & \text{sPHENIX, at } (y_3^*,y_4^*)\simeq(0.44,\,1.1),\\[2pt]
0.73 & \text{ATLAS/CMS, at } (y_3^*,y_4^*)\simeq(0.67,\,2.5),\\[2pt]
0.91 & \text{LHCb, at the corner } (y_3^*,y_4^*)=(2.0,\,5.0),
\end{cases}
\end{equation}
the sPHENIX value matching the cutoff visible in
Fig.~\ref{fig:helac_comparison}.  Two trends are apparent: increasing
$p_T^{\min}$ at fixed acceptance shrinks the accessible $\cos\chi$
monotonically toward $-1$, while a wider or more forward acceptance
enlarges it toward $+1$---so the forward LHCb window reaches the largest
$\cos\chi_{\max}^{\rm fid}$ of the three.
 
\bibliography{references}
 
\end{document}